\documentclass[sigconf]{acmart}
\renewcommand\footnotetextcopyrightpermission[1]{} 

\AtBeginDocument{%
  \providecommand\BibTeX{{%
    \normalfont B\kern-0.5em{\scshape i\kern-0.25em b}\kern-0.8em\TeX}}}

\copyrightyear{2023}
\acmYear{2023}
\setcopyright{rightsretained}
\acmConference[MM '23]{Proceedings of the 31st ACM International Conference on Multimedia}{October 29-November 3, 2023}{Ottawa, ON, Canada}
\acmBooktitle{Proceedings of the 31st ACM International Conference on Multimedia (MM '23), October 29-November 3, 2023, Ottawa, ON, Canada}
\acmDOI{10.1145/3581783.3612091}
\acmISBN{979-8-4007-0108-5/23/10}

\usepackage{graphicx}

\usepackage{algorithm, algpseudocode}
\usepackage{multirow}
\usepackage{caption}
\usepackage{comment}
\usepackage{color}
\usepackage{enumitem}

\definecolor{citecolor}{RGB}{119,185,0} 

\usepackage{pifont}

\usepackage{amssymb}
\usepackage{amsmath}
\usepackage{animate}
\usepackage{graphicx}

\usepackage{booktabs}
\usepackage{multirow}
\usepackage{makecell}
\usepackage{adjustbox}
\usepackage{bm}
\usepackage{hyperref}
\usepackage{balance}

\newlength\savewidth

\def\eg{\emph{e.g.}}

\hypersetup{
    colorlinks=true,
    linkcolor=blue,
    urlcolor=blue,
    citecolor=blue
}

\copyrightyear{2023}
\acmYear{2023}
\setcopyright{rightsretained}
\acmConference[MM '23]{Proceedings of the 31st ACM International Conference on Multimedia}{October 29-November 3, 2023}{Ottawa, ON, Canada}
\acmBooktitle{Proceedings of the 31st ACM International Conference on Multimedia (MM '23), October 29-November 3, 2023, Ottawa, ON, Canada}
\acmDOI{10.1145/3581783.3612091}
\acmISBN{979-8-4007-0108-5/23/10}

\makeatletter
\gdef\@copyrightpermission{
 \begin{minipage}{0.3\columnwidth}
  \href{https://creativecommons.org/licenses/by/4.0/}{\includegraphics[width=0.90\textwidth]{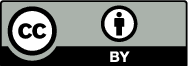}}
 \end{minipage}\hfill
 \begin{minipage}{0.7\columnwidth}
  \href{https://creativecommons.org/licenses/by/4.0/}{This work is licensed under a Creative Commons Attribution International 4.0 License.}
 \end{minipage}
 \vspace{5pt}
}
\makeatother

\begin{document}
\title{Online Distillation-enhanced Multi-modal Transformer \\ for Sequential Recommendation}

\author{Wei Ji}
\authornote{Equal Contribution}
\affiliation{%
  \institution{National University of Singapore}
  \city{Singapore}
  \country{Singapore}
}
\email{weiji0523@gmail.com}

\author{Xiangyan Liu}
\authornotemark[1]
\affiliation{%
  \institution{National University of Singapore}
  \city{Singapore}
  \country{Singapore}
}
\email{liu.xiangyan@u.nus.edu}

\author{An Zhang}
\authornote{Corresponding Author}
\affiliation{%
  \institution{National University of Singapore}
  \city{Singapore}
  \country{Singapore}
}
\email{anzhang@u.nus.edu}

\author{Yinwei Wei}
\affiliation{%
  \institution{Monash University}
  \city{Melbourne}
  \country{Australia}
}
\email{weiyinwei@hotmail.com}

\author{Yongxin Ni}
\affiliation{%
  \institution{National University of Singapore}
  \city{Singapore}
  \country{Singapore}
}
\email{niyongxin@u.nus.edu}

\author{Xiang Wang}
\authornote{Xiang Wang is also affiliated with Institute of Artificial Intelligence, Institute of Dataspace, Hefei Comprehensive National Science Center}
\affiliation{%
\institution{University of Science and Technology of China}
 \city{Hefei}
 \country{China}}
\email{xiangwang1223@gmail.com}

\renewcommand{\shortauthors}{Wei Ji et al.}

\begin{abstract}
Multi-modal recommendation systems, which integrate diverse types of information, have gained widespread attention in recent years. However, compared to traditional collaborative filtering-based multi-modal recommendation systems, research on multi-modal sequential recommendation is still in its nascent stages. Unlike traditional sequential recommendation models that solely rely on item identifier (ID) information and focus on network structure design, multi-modal recommendation models need to emphasize item representation learning and the fusion of heterogeneous data sources. This paper investigates the impact of item representation learning on downstream recommendation tasks and examines the disparities in information fusion at different stages. Empirical experiments are conducted to demonstrate the need to design a framework suitable for collaborative learning and fusion of diverse information. Based on this, we propose a new model-agnostic framework for multi-modal sequential recommendation tasks, called \textbf{Online Distillation-enhanced Multi-modal Transformer (ODMT)}, to enhance feature interaction and mutual learning among multi-source input (ID, text, and image), while avoiding conflicts among different features during training, thereby improving recommendation accuracy. To be specific, we first introduce an ID-aware Multi-modal Transformer module in the item representation learning stage to facilitate information interaction among different features. Secondly, we employ an online distillation training strategy in the prediction optimization stage to make multi-source data learn from each other and improve prediction robustness. Experimental results on a stream media recommendation dataset and three e-commerce recommendation datasets demonstrate the effectiveness of the proposed two modules, which is approximately 10\% improvement in performance compared to baseline models. Our code will be released at: 
\url{https://github.com/xyliugo/ODMT}.

\end{abstract}



\maketitle

\vspace{-0.16cm}
\section{Introduction}

With the emergency of multimedia platforms (\eg, TikTok, Youtube), multi-modal recommendation system is becoming increasingly important in both academic \cite{wei2023multi, zhou2023enhancing, zhang2023multimodal,wei2022causal} and industry \cite{xu2022rethinking, baltescu2022itemsage}.
It aims to understand user preferences of multi-modal content information, based on their historical behaviors (\eg, clicks, comments). 

Compared to general recommendation systems, multi-modal recommendation requires not only model architecture design, but also consideration of how to effectively apply multi-modal features in downstream tasks, especially in a way compatible with the current recommendation system. Most current recommendation systems rely on collaborative filtering \cite{he2017neural, wang2019neural, he2020lightgcn,BC-loss,InvCF}, which predicts user-item interactions by modeling users and items separately and then computing the similarity between the user and candidate item to generate a prediction score. In general multi-modal recommendation systems, multi-modal features are used to enhance the connection between user-item pairs \cite{wei2019mmgcn, wei2021hierarchical, wei2023multi} or as side information, which is complementary to ID features \cite{he2016vbpr, zhang2019feature}. In contrast, sequential recommendation systems rely more on item representation learning than collaborative filtering and model user representations according to items clicked by the user and their temporal sequences \cite{kang2018self, sun2019bert4rec, wu2021mm, zhou2020s3, wu2019session}. This approach places a higher demand on item representation learning, especially in the case of multi-modal recommendation systems, where the raw item information is more diverse.

Our study focuses on item representation learning in multi-modal sequential recommendation systems and explores the performance of single-modal features (text or image) as individual input and as input combined with ID features in downstream recommendation tasks. We also investigate different fusion strategies when combining ID and multi-modal information. Our exploration experiments reveal that compared to other network structures, Transformers provide better semantic transformation and representation learning for single-modal features, leading to more accurate predictions in recommendation tasks. However, the advantage of strong representation brought by Transformers weakens when using multi-source data (ID, text, and image) as input. Further analysis reveals that this is due to ID features being easier to optimize and producing lower training loss in recommendation prediction tasks, whereas multi-modal features provide valuable prior information on item similarity, making recommendation systems easy to retrieve items of interest for users. Therefore, when ID and modal information are combined, the improvement in evaluation metrics may not perfectly align with the direction of the loss reduction.

To address the challenges in multi-modal sequential recommendation, we propose a new model-agnostic framework called Online Distillation-enhanced Multi-modal Transformer (ODMT) equipped with two novel modules. Firstly, we introduce an ID-aware Multi-modal Transformer module in the item representation learning stage to facilitate information interaction among different features. Secondly, we apply an online distillation training strategy in the prediction optimization stage to obtain more robust predictions without compromising the loss optimization of the multi-modal features. Overall, our contributions can be summarized as follows:

\begin{itemize}[itemsep=2pt,topsep=1mm,parsep=0pt,leftmargin=5.5mm]

    \item To relieve the incompatibility issue between multi-modal features and existing sequential recommendation models, we introduce the ODMT framework, which comprises an ID-aware Multi-modal Transformer module for item representation.
    
    \item To obtain robust predictions from multi-source input, we propose an online distillation training strategy in the prediction optimization stage, which marks the first instance of applying online distillation to a multi-modal recommendation task.

    \item Comprehensive experiments on \textbf{four} diverse multi-modal recommendation datasets and \textbf{three} popular backbones for sequential recommendation validate the effectiveness and transferability of proposed method, which is about 10\% performance improvement compared with other baseline models.
\end{itemize}

\section{PRELIMINARIES}
This section aims to thoroughly investigate the effects of Item Representation Learning (IRL) and Information Fusion (IF) modules on downstream recommendation networks.

We provide empirical evidence through experiments, which highlight two key findings:  1) Transformers are effective in transforming multi-modal information from general semantic to specific recommendation semantic; 2) simple fusion strategies can cause a discrepancy between the direction of loss optimization and the direction of metric improvement during the training process, thus affecting the significance of multi-modal features in the recommendation model.

\subsection{Brief Concepts}
The IRL module is responsible for generating the final item embeddings by converting raw input data into distributed representations. Input data can be categorized as item ID or item other modalities (\eg, image and text). The embedding table of items is a crucial component in sequential recommendation models~\cite{kang2018self, hidasi2015session}. Each item has a unique embedding representation that corresponds to its index. For multi-modal data, BERT~\cite{devlin2018bert} and ViT~\cite{dosovitskiy2020image} are utilized to extract textual and visual features from raw data. The extracted multi-modal features are then fed into a Feature Semantic Transformation (FST) module to convert modal information into a semantic space suitable for recommendations. Our FST module candidates include DNNs, MoE Adaptor (MoE), and Transformers+DNNs (TRM+DNNs), which have been widely used in previous researches~\cite{liu2022disentangled, hou2022towards, wang2022transrec, yuan2023go}.

In multi-modal sequential recommendation models, the IF module can be categorized into Early Fusion and Late Fusion \cite{hou2022learning, rashed2022carca, zhang2019feature, pan2022multimodal}. Early Fusion involves embedding all item information into a single feature representation, which is then inputted into the model, while Late Fusion is based on the prediction results or the prediction scores of each feature. In this paper, we consider three types of input information, namely ID, text, and image, that need to be fused. In Early Fusion, we obtain fused item embeddings by averaging the three features. In contrast, in Late Fusion, we get a user's general preferences by averaging the three logits, where different logits correspond to different user preferences. Intermediate Fusion is not discussed in this section, as it is a model-dependent fusion method \cite{zhou2023comprehensive}.

\begin{figure}[t]
    \begin{minipage}[t]{\linewidth}
      \centering
      \includegraphics[width=\linewidth]{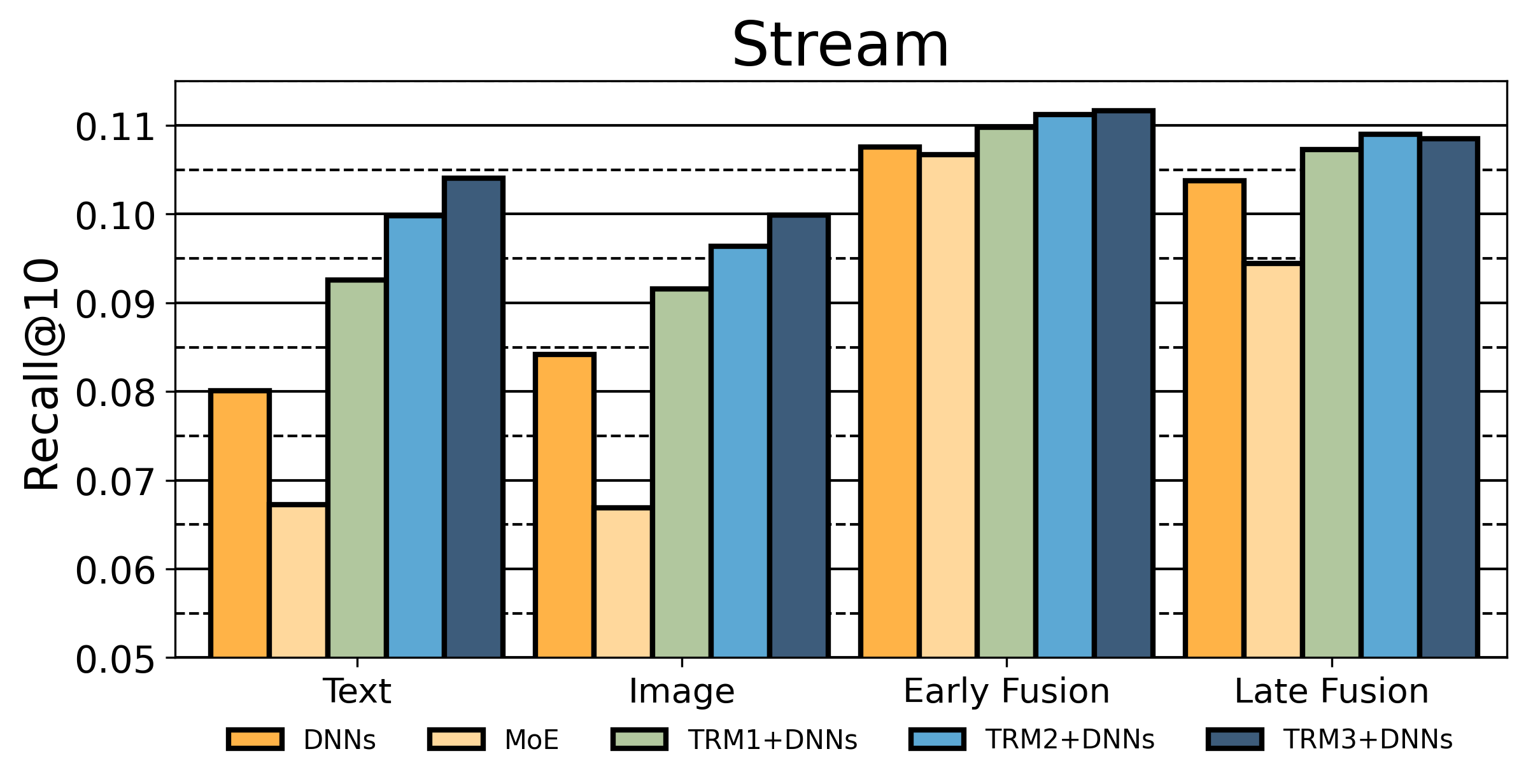}
    \end{minipage}
    \begin{minipage}[t]{\linewidth}
      \centering
      \includegraphics[width=\linewidth]{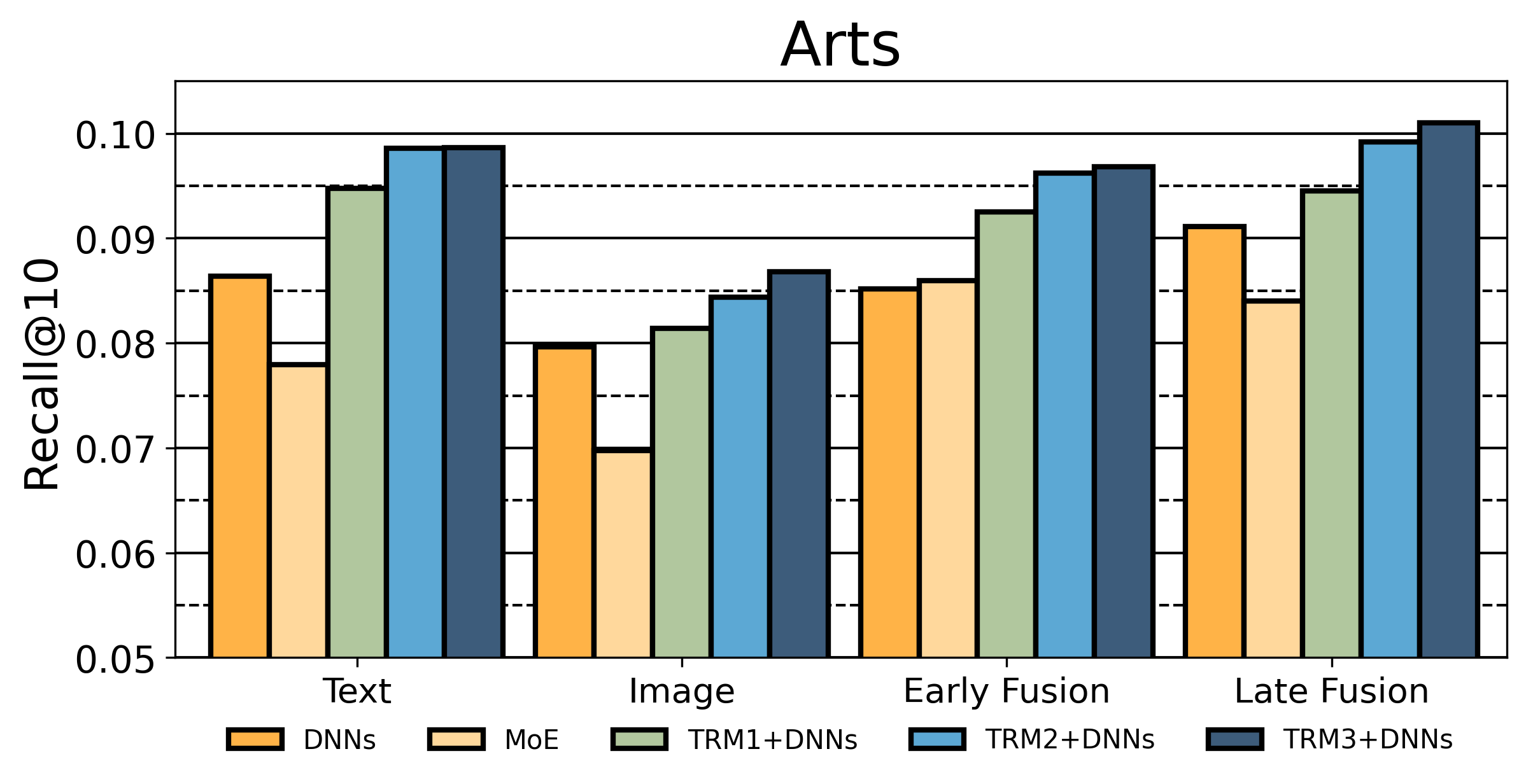}
    \end{minipage}
\vspace{-0.8cm}
  \caption{Comparison of Recall@10 for different FST modules based on single-modal and multi-source input, across Arts and Stream datasets. "x" in TRMx+DNNs represents the number of Transformer layers. For x = 1, it represents 1 Transformer layer for text and 1 Transformer layer for image.}
  \label{exp:exploration1}
\end{figure}

\begin{figure*}[h]
    \begin{minipage}{0.475\columnwidth}
      \includegraphics[width=\linewidth]{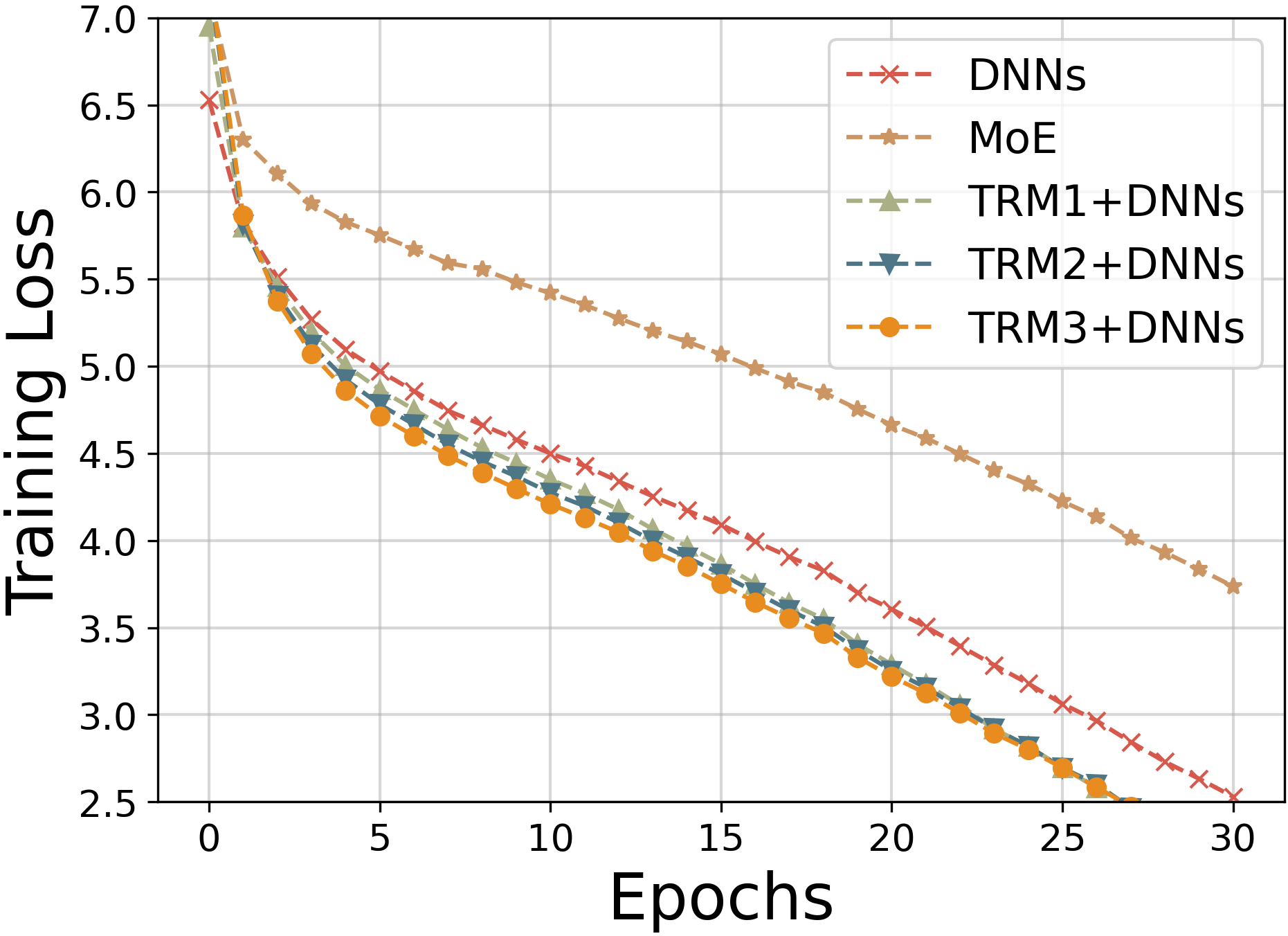}
       (a) Training Loss v.s. FST
    \end{minipage}
    \begin{minipage}{0.475\columnwidth}
      \includegraphics[width=\linewidth]{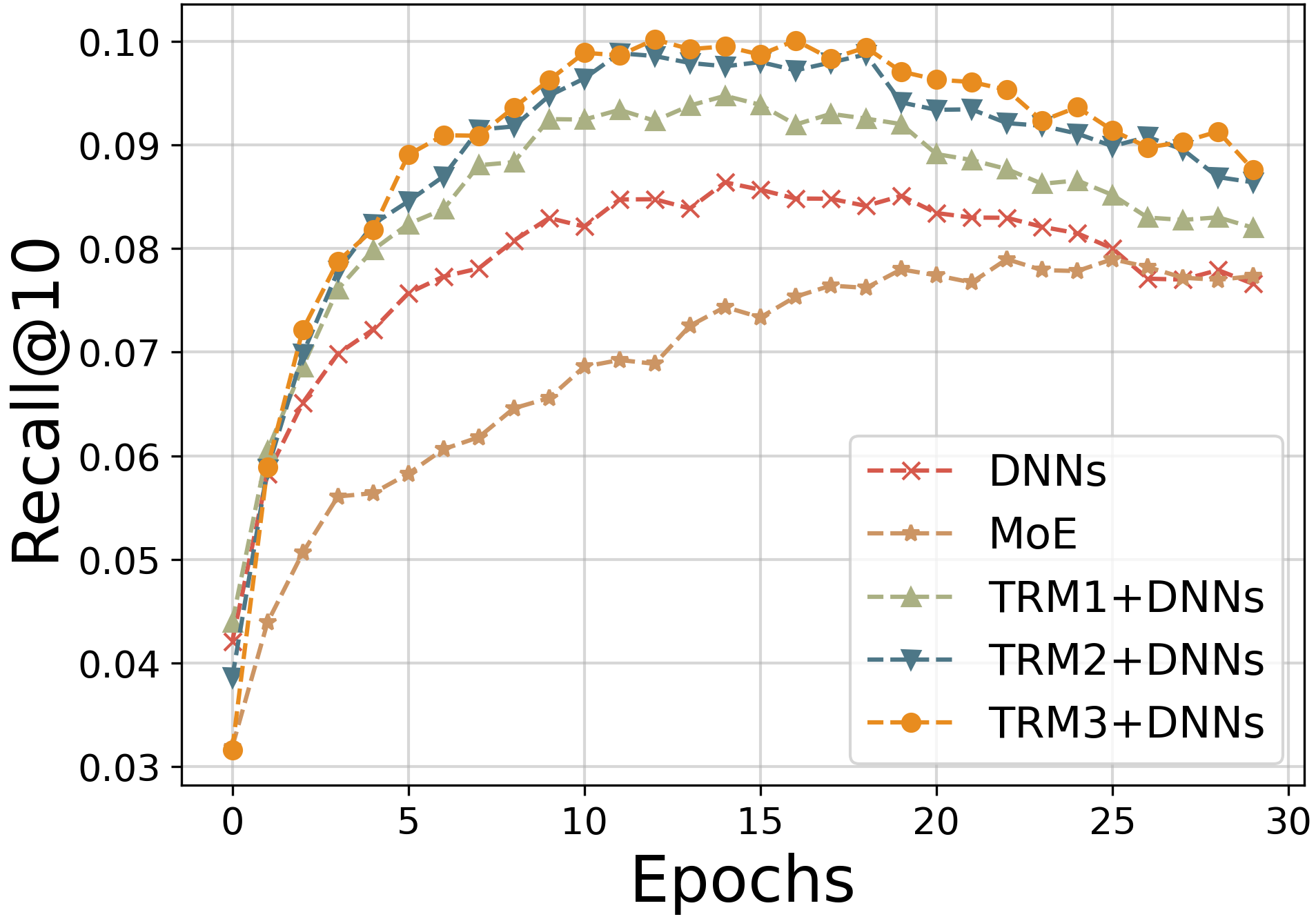}
       (b) Testing Recall v.s. FST
    \end{minipage}
    \begin{minipage}{0.475\columnwidth}
      \includegraphics[width=\linewidth]{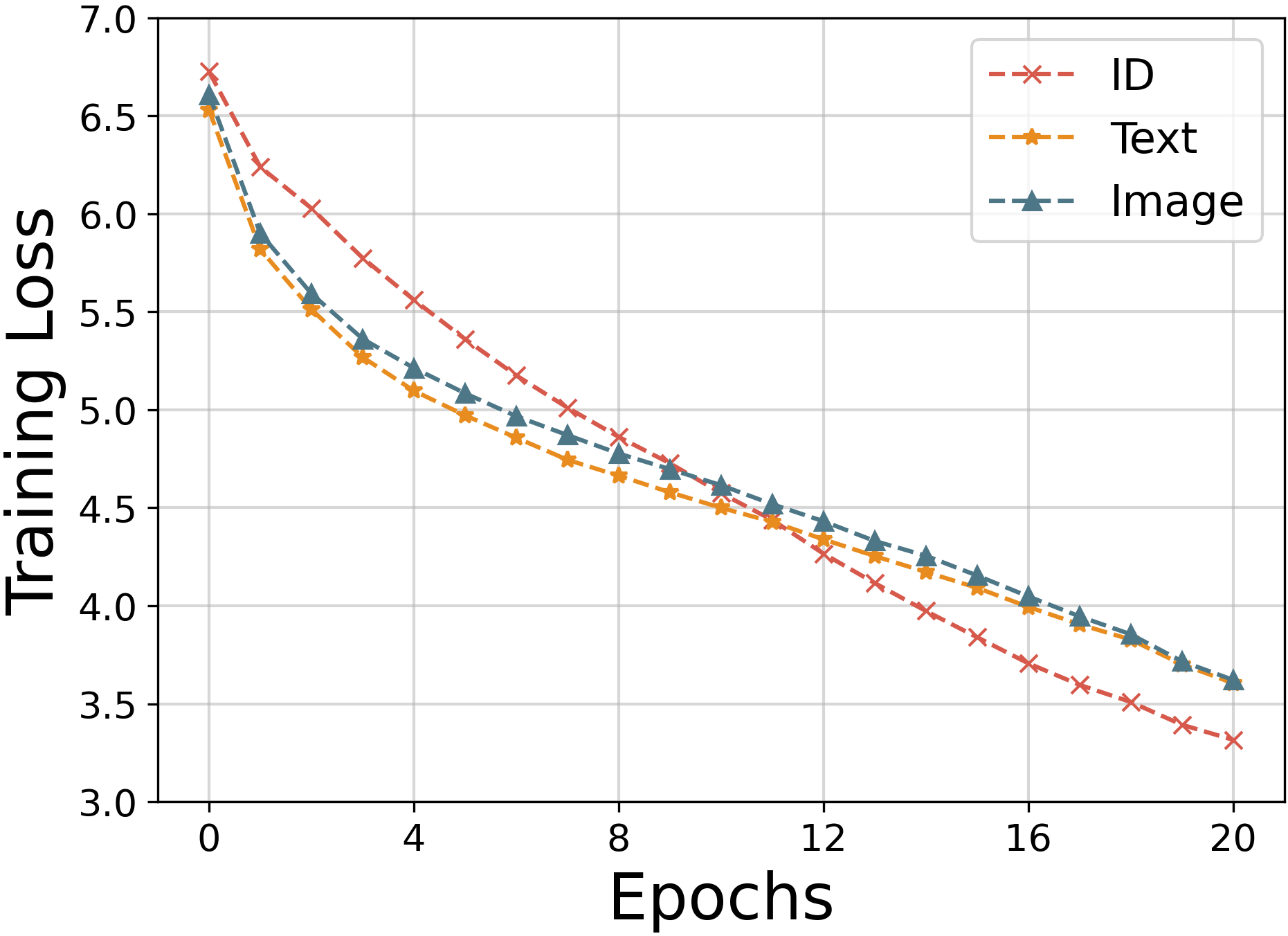}
       (c) Training Loss v.s. Input
    \end{minipage}
    \begin{minipage}{0.475\columnwidth}
      \includegraphics[width=\linewidth]{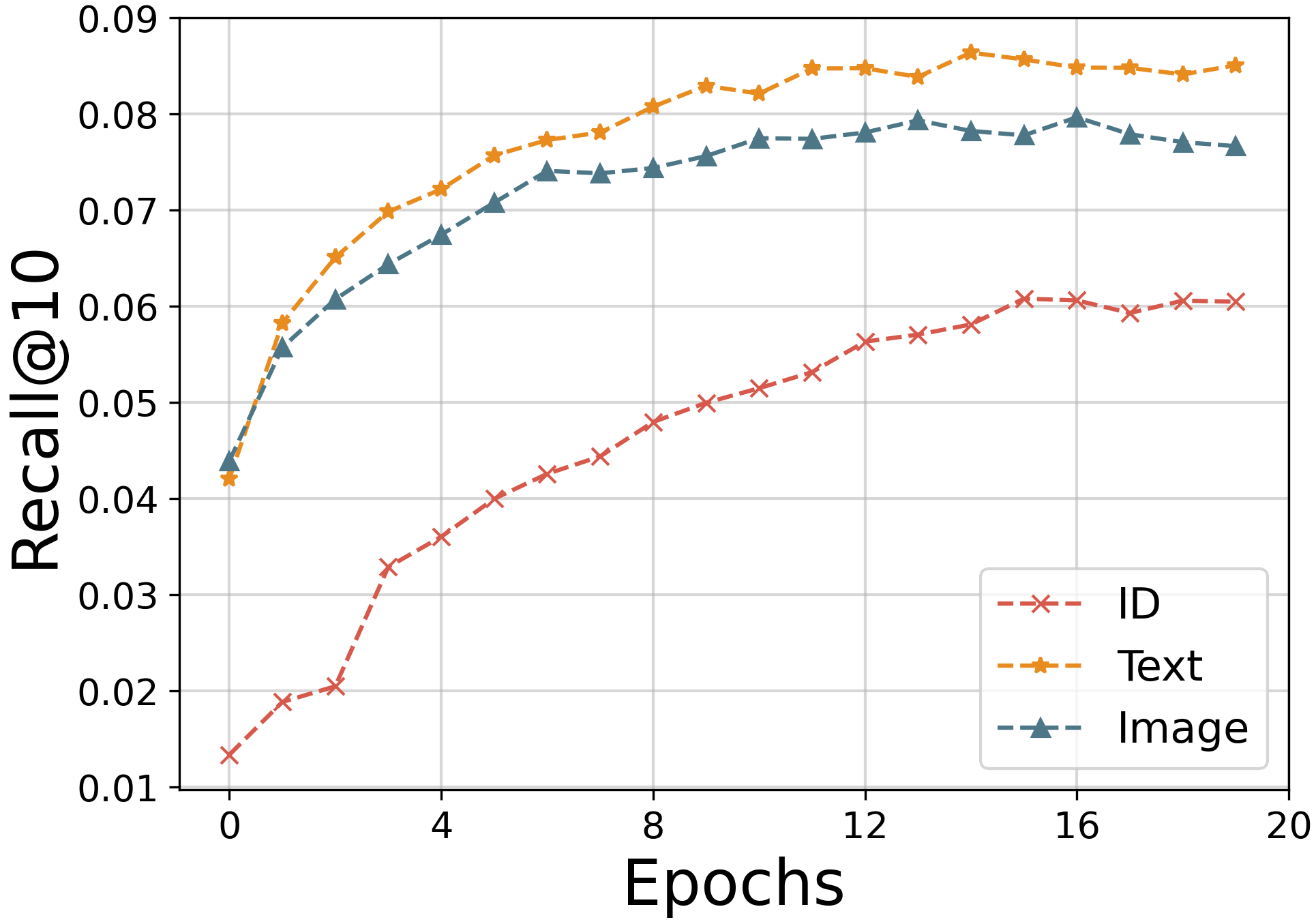}
       (d) Testing Recall v.s. Input
    \end{minipage}
    \vspace{-0.3cm}
    \caption{Training curves of FST modules with text input, evaluating (a) training loss and (b) testing Recall@10, and training curves of different single input with DNNs as FST module, evaluating (c) training loss and (d) testing Recall@10 on the Arts dataset.}
    \label{exp:exploration2}
\end{figure*}

\subsection{Empirical Explorations}
To ensure fair comparison in our experiments, we control all variables other than the FST module and IF module, including random seed, pre-trained encoders, hyper-parameters (\eg,  learning rate, embedding size, hidden size, and dropout ratio), and experimental codes, all experiments are conducted in a unified framework. The backbone model for our sequential model is the representative one, SASRec ~\cite{kang2018self}, which uses self-attention mechanisms for sequence modeling. Figure ~\ref{exp:exploration1} shows experimental results on the Stream and Arts (Amazon) datasets.

General visual and textual features extracted by pre-trained models (\eg, BERT and ViT) are not necessarily suitable for recommendation tasks. Therefore, FST module is needed to transform the modality features into recommendation semantics. Figure \ref{exp:exploration1} indicates that Transformers are capable of performing semantic transformation more effectively than DNNs and MoE when inputted with single-modal data. This finding highlights the potential of Transformers in learning powerful representations for recommendation systems.

Previous studies in multi-modal sequential recommendation models ~\cite{zhang2019feature, hou2022towards, pan2022multimodal, rashed2022carca} have typically used simple FST modules, such as DNNs and MoE, with Early Fusion or Late Fusion methods. In these cases, fusing multiple information sources has generally yielded favorable outcomes. When Transformers are employed as the FST module, we notice a diminishing advantage of utilizing multi-source input instead of single-modal input as the number of Transformer layers increases. Remarkably, even on the Arts dataset, the effectiveness of the single-modal input surpasses that of the multi-source fusion input. These results suggest that while Transformers possess strong representation learning capabilities, they may not be able to fully showcase their potential in the scenarios of multi-source input.

To provide more insights into the impact of FST modules and IF modules, we fix the input with plain text and the FST module with DNNs, respectively. Figure ~\ref{exp:exploration2}a and ~\ref{exp:exploration2}b illustrate the consistency between training loss and testing Recall, revealing that models with lower training loss corresponded to higher Recall scores. This observation suggests that single-modal input enables the FST module to better learn item representations and effectively reduce the training loss, leading to improved downstream recommendation performance. On the other hand, Figure ~\ref{exp:exploration2}c and \ref{exp:exploration2}d  demonstrate the inconsistency between training loss and testing Recall when different types of information are used as input. This inconsistency may arise from overfitting or the mismatch between the objective function and the evaluation metric. For recommendation models, optimizing the ID features can be viewed as an unconstrained optimization problem, resulting in lower training loss. Conversely, optimizing modality features is subject to constraints due to the prior information that items' content has similarities. Therefore, even if the modality-based models do not achieve significantly lower training loss, they can achieve better performance, particularly when using Transformers as the FST module.

Based on our findings, we can conclude that the FST module plays a crucial role in extracting informative representations of items. Integrating multi-modal information  may not lead to the optimal improvement in recommendation metrics, as there exists a misalignment between the direction of metric improvement and loss reduction. This misalignment creates a dilemma in learning both ID and modality features simultaneously for recommendation systems. The dilemma becomes more challenging as the modality representation capacity increases, which can ultimately lead to compromised recommendation performance. And in some cases, the performance is even worse compared to single-modal models. 

\section{Method}
\begin{figure*}
\centering
\includegraphics[width=\linewidth]{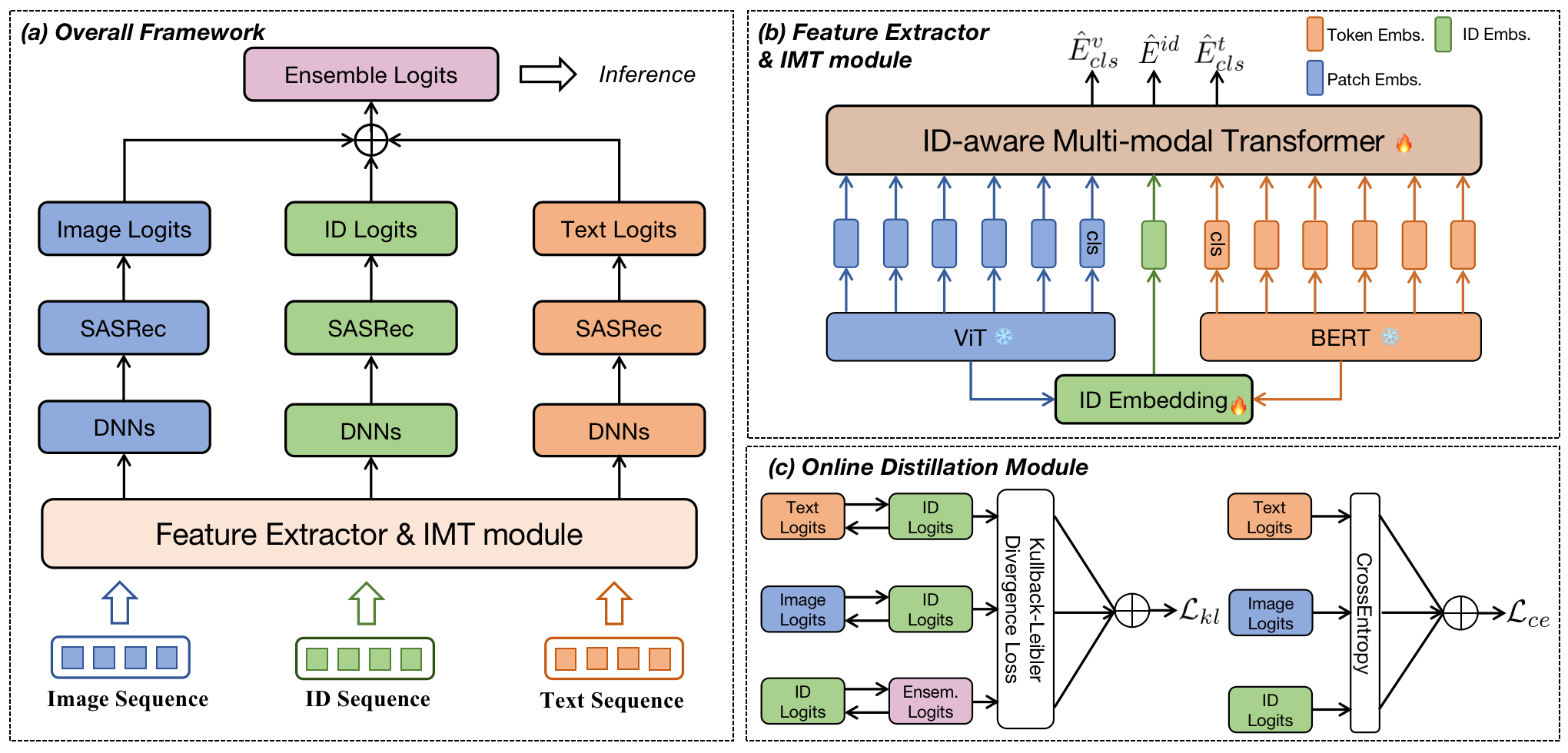}
\vspace{-0.8cm}
\caption{ (a) Overall framework of our proposed ODMT model, which illustrates the forward computation flow based on modifications to the Late Fusion approach. (b) shows our proposed IMT module, and (c) represents our proposed Online Distillation module. Different colors represent the information flow of different features (\textcolor[RGB]{169,209,142}{ID}, \textcolor[RGB]{143,170,220}{image}, and \textcolor[RGB]{244,177,131}{text}).}
\label{exp:model}
\end{figure*}

In the previous section, we discussed the significance of Transformers in acquiring multi-modal representations and also pointed out the shortcomings of current methods when fusing multi-source information. To further improve the representation ability of Transformers in recommendation scenarios and enable collaborative learning from different modalities without conflicts during training, we propose two modules based on the Late Fusion framework: 1) \textbf{ID-aware Multi-modal Transformer}. We incorporate the ID features with modality features and perform fine-grained feature interactions within a single multi-modal Transformer. 2) \textbf{Online Distillation}. We use an online distillation framework to compute the recommendation classification loss for each input, leveraging the strong representation capacity of Transformers. This ensures that each sub-network captures distinct user preferences by optimizing corresponding loss. Besides, we introduce a distillation loss that facilitates on-the-fly mutual learning~\cite{zhang2018deep, guo2020online} among the student networks. Figure \ref{exp:model} shows the overall framework of ODMT with the above two modules.

\subsection{Notations}

We define the set of users as $\mathcal{U}=\{u\}$ and the set of items as $\mathcal{A}=\{a\}$. For the each item $a_j$, we record its image, text and ID as $a_j^{v}$, $a_j^{t}$, and $a_j^{id}$, respectively. The $i$-th user interaction sequence $S_i$, is defined in chronological order as $S_i^m=\{a_{s_1}^m(i), a_{s_2}^m(i),..., \}$, where $a_{s_k}^m(i)$ represents the $k$-th interaction item, and $m$ represents the types of the item, i.e., image, text, and ID.

\subsection{Item Representation Learning}
\textbf{Feature Extractor.} Given an item with different types ($a^{v}$, $a^{t}$, and $a^{id}$), we first use fixed visual and textual feature extractors (ViT and BERT) to obtain the corresponding fine-grained patch-level and token-level features, then we obtain the corresponding ID features from a learnable embedding table, the feature extraction process is summarized as follows:
\begin{equation}
\begin{aligned}
E^{v}={\rm ViT}(a^{v}),\ E^{t}={\rm BERT}(a^{t}),\ E^{id}={\rm EmbeddingTable}(a^{id})
\end{aligned}
\end{equation}
where $E^{v}=[E^{v}_1;...;E^{v}_{n_{v}}; E^{v}_{cls}]\in \mathbf{R}^{d_{v}\times (n_{v}+1)}$, $E^t=[E^t_{cls};E^{t}_1;...;E^{t}_{n_{t}}]\in \mathbf{R}^{d_{t}\times (n_{t}+1)} $ and $E^{id}\in\mathbf{R}^{d_{id}}$. $d_v$, $d_t$, and $d_{id}$ are the visual feature dimension, textual feature dimension, and ID feature dimension respectively. $n_v$ is the number of image patches and $n_t$  is the number of word tokens. $E_{cls}$ here represents the embedding of the special token "[CLS]". 

Afterward, we use a simple feature transformation matrix to project each input feature into the same dimension $d$ as $\tilde {E}^m=W^m \cdot E^m+b^m$, where $W^m\in\mathbf{R}^{d\times d_m}$, $b^m\in\mathbf{R}^{d}$, and $m\in\{v,t,id\}$.

\textbf{ID-aware Multi-modal Transformer (IMT).} In this part, we describe the proposed ID-aware Multi-modal Transformer (IMT) module, which consists of multiple standard Transformer layers. Different from traditional multi-modal Transformers designed for visual and textual features, our IMT module integrates the unique ID features in the recommendation system into Transformers. Our goal is to obtain a unified framework that transforms item embeddings from the original generic feature space to one that is suitable for recommendations (especially for modality features).

To achieve this, we first concatenate visual patch features,  ID features, and textual token features  together as $\tilde{E}=[\tilde{E}^v;\tilde{E}^{id};\tilde{E}^t]\in\mathbf{R}^{d\times(n_v+n_t+3)}$, where $[\cdot]$ denotes the concatenation operation. Since there are no paddings for visual and ID features, we set the mask value for all visual and ID features to 0 and obtain the attention mask as $\tilde{M}\in \mathbf{R}^{(n_v+n_t+3)\times(n_v+n_t+3)}$. However, in the previous section of the discussion, we discovered that the ID features could influence the optimization direction of the model. To prevent the misleading influence of ID embeddings on modality embeddings, we make the following adjustments to the original attention mask matrix as $\tilde{M}[:n_v+1, n_v+1]=1$, $\tilde{M}[n_t+2:,n_v+1]=1$. This ensures that the ID embeddings can attend to the modality embeddings, while the modality embeddings cannot attend to the ID embeddings

Similar to the traditional Transformer modeling process,  once we have input feature $\tilde{E}$ and revised attention mask matrix $\tilde{M}$, we can feed them directly into the standard Transformer layer as:
\begin{equation}
\begin{aligned}
\hat{E}={\rm IMT}(\tilde{E},\tilde{M};{\rm \Theta_{IMT}})
\end{aligned}
\end{equation}
where $\Theta_{{\rm IMT}}$ denotes all the learnable parameters in the IMT module and $\hat{E}=[\hat{E}^v;\hat{E}^{id};\hat{E}^t]\in \mathbf{R}^{d\times(n_v+n_t+3)}$  denotes the encoded item representation. Then we use the "cls" embedding to represent the global feature for image and text. We do not need to include additional positional embeddings in the input embeddings, as the features extracted from the pre-trained models already contain positional information.

To obtain more powerful feature representations for the recommendation domain \cite{liu2022disentangled}, we empirically employ separate two-layer DNNs with a LeakyRelu activation layer \cite{xu2015empirical} for each output as:
\begin{equation}
\begin{aligned}
{D}^k&=W^k_2\cdot{\rm LeakyRelu}(W^k_1\cdot \hat{E}^m_{cls}+b^k_1)+b^k_2 \\ 
{D}^{id}&=W^{id}_2\cdot{\rm LeakyRelu}(W^{id}_1\cdot \hat{E}^{id}+b^{id}_1)+b^{id}_2
\end{aligned}
\end{equation}
where $W^k_i, W^{id}_i\in \mathbf{R}^{d\times d}$, $k\in \{v,t\}$ and $i\in \{1,2\}$. ${D}$ represents the final embedding of the item. In detail, ${D}^{m}_k$ represents the $k$-th item embedding whose input is $m$, where $m\in\{v,t,id\}$.

\textbf{ID Embedding Initialization.} It is noteworthy that the initial embedding table for the ID features is typically randomly generated, which differs significantly from the text and image features extracted by large pre-trained models such as BERT \cite{devlin2018bert} and ViT \cite{dosovitskiy2020image}. In the IMT module, a self-attention mechanism is utilized to compute the similarity between queries and keys. The discrepancy between the ID features and modality features can negatively impact the optimization of the IMT module. To address this, when $d_v=d_t$, we initialize the ID embedding table by averaging $E^v_{cls}$ and $E^t_{cls}$, thus establishing $d_v=d_t=d_{id}$. When $d_v\neq d_t$, the ID embedding table is initialized using either the text or image features or either the concatenated text and image features, arbitrarily chosen.

\subsection{User Sequence Modeling} 
In sequential recommendation tasks, user sequence features are generated from interacted items. The widely used SASRec \cite{kang2018self} model employs a multi-head attention mechanism for user sequence modeling. We adopt SASRec as the backbone network to learn user sequence representations from three input types (image, text, and ID). The user behavior sequence with final item embeddings is represented as $\mathbf{S}^m=\{{D}^m_{s_1},{D}^m_{s_2},...,{D}^m_{s_n}\}$. Using this sequence, we obtain the user sequence feature $H^m$ as follows:
\begin{equation}
\begin{aligned}
H^m={\rm {SASRec}}_m(\mathbf{S}^m;\Theta_{{\rm SASRec}_m})
\end{aligned}
\end{equation}
where $H^{m}\in\mathbf{R}^{d}$ is the user preference feature, $\Theta_{{\rm SASRec}_m}$ denotes all the learnable parameters in ${\rm SASRec}_m$, $m$ represents the input type.

\subsection{Debiased Inbatch Loss}
Following prior research \cite{zhang2019feature, hou2022towards, yuan2023go}, we adopt next-item prediction as the recommendation task, with negative log-softmax loss as the guiding loss function, which helps to bring user preference and the target item closer in the feature space. Given the user interaction sequence as $a_1 \rightarrow a_n$, the positive sample that needs to be predicted is $a_{n+1}$. 

For computational efficiency, we utilize all the items from the user interaction sequences in the mini-batch as the candidate item sets. However, this approach leads to a distribution of items in the candidate item sets, known as the Matthew effect, where the majority of items are highly popular with a large number of interactions, causing popular items to become over-represented and leading to under-optimized performance for less popular items. To mitigate this effect, we debias the similarity computation results between users and items based on popularity \cite{yi2019sampling, chen2022cache}.

It is noteworthy to consider false negatives in in-batch sampling. When using items that the user has already interacted with as negative samples, the gradient descent direction of the model may be confused. To address this issue, items in the candidate item set that overlap with the user's interaction sequence should be excluded when predicting the to-click item of a user.

In the batch training process, a set of $B$ training instances is considered, where each instance corresponds to a user sequence and a target next item (positive sample). These instances are encoded as embedding representations $\{(H_{h_1},{D}_{d_1}),(H_{h_2},{D}_{d_2}),...,(H_{h_B},{D}_{d_B})\}$, where $h_i$ and $d_i$ represent the indices of the user sequence and target item of the $i$-th pair, respectively. Then we define the candidate item sets which exclude the items that overlap with $h_i$-th user sequence as $\mathcal{B}_{h_i}$. Finally, we adopt the cross-entropy loss as the objective function:
\begin{gather}
\mathcal{L}_{ce}=-\sum_{i=1}^B\log \frac{\exp(s(H_{h_i},D_{d_i}))}{\exp(s(H_{h_i},D_{d_i}))+\sum_{d_j\in \mathcal{B}_{h_i}}\exp(s(H_{h_i},{D}_{d_j}))}\\
s(H_{h_i},D_{d_i})=H_{h_i} \cdot {D}_{d_i} - {\rm log}(pop(a_{d_i}))
\end{gather}
where $pop(a_{d_i})$ represents the frequency of item $a_{d_i}$ appearing in the training set.

\subsection{Online Distillation}
Similar to Late Fusion, we model different types of user sequences to calculate the similarity between user sequences and the target items as well as candidate items, obtaining logits that represent the user's interest distribution. However, different from Late Fusion, we treat each part as a student network branch and directly calculate the classification loss for each corresponding logit, rather than averaging multi-source logits and obtaining a classification loss. We believe that this independent loss calculation approach will alleviate conflicts between multiple features during the training process. 
In detail, we denote ${\rm \mathbf{z}^m}$ and  ${\rm \mathbf{y}}$ as the logits and ground truth, where $m\in\{v,t,id\}$. Late Fusion method obtains the final classification loss as $\mathcal{L}_{ce}={\rm cross\_entropy}((\mathbf{z}^{e},\mathbf{y})$, where $\mathbf{z}^{e}$ is the ensemble logits as $\mathbf{z}^{e}=\frac{\mathbf{z}^{v}+\mathbf{z}^{t}+\mathbf{z}^{id}}{3}$ . As for collaborative learning, we calculate classification loss as follows:
\begin{gather}
\mathcal{L}_{ce}=\mathcal{L}^{v}_{ce}+\mathcal{L}^{t}_{ce}+\mathcal{L}^{id}_{ce}\\
\mathcal{L}^{m}_{ce}={\rm cross\_entropy}(\mathbf{z}^{m},\mathbf{y}),\ m\in\{v,t,id\}
\end{gather}
In the knowledge distillation part, we calculate the distillation loss as follows:
\begin{equation}
\begin{aligned}
\mathcal{L}_{kl}&=\mathcal{L}^{v}_{kl}+\mathcal{L}^{t}_{kl}+\mathcal{L}^{id}_{kl}\\
\mathcal{L}^{v}_{kl}&={\rm T}^2KL(\sigma(\mathbf{z}^{id}/{\rm T}),\sigma(\mathbf{z}^{v}/{\rm T})) \\
\mathcal{L}^{t}_{kl}&={\rm T}^2KL(\sigma(\mathbf{z}^{id}/{\rm T}),\sigma(\mathbf{z}^{t}/{\rm T})) \\
\mathcal{L}^{id}_{kl}&={\rm T}^2KL(\sigma(\mathbf{z}^{e}/{\rm T}),\sigma(\mathbf{z}^{id}/{\rm T}))\\
\end{aligned}
\end{equation}
where {\rm T} is the temperature parameter, $\sigma$ is the softmax operation, and $KL(p,q)$ means the KL divergence between the soften outputs $p$ from teacher network and $q$ from student network.

Because at the beginning of the model training, the predictions of each student network are not accurate enough, we need to decrease the weight of the distillation loss during the early training stage. Therefore, we adopt a time-dependent unsupervised ramp-up function $w(\alpha)$\cite{laine2016temporal}. When the training epoch is 0, $w(\alpha)$ is 0. Then, $w(\alpha)$ increases exponentially as the training epoch progresses. When the training epoch reaches $\alpha$, $w(\alpha)$ takes a value of 1. Then the final total loss is as follows:
\begin{gather}
\mathcal{L}_{total}=\mathcal{L}_{ce}+w(\alpha)\cdot\mathcal{L}_{kl}
\end{gather}

\begin{table}[ht]
    \centering
    \caption{Statistics of all datasets used in our experiment. "\#Inter." represents total user-item interactions, and "Avg. $u$" denotes the average user length.}\label{exp:dataset}
    \vspace{-0.4cm}
    \begin{tabular}{*{6}{r}}
    \toprule
        \textbf{Dataset} & \textbf{\#Users} & \textbf{\#Items} & \textbf{\#Inter.} & \textbf{Avg. $u$} & \textbf{Density} \\ \midrule
        \textbf{Stream} & 100,000 & 19,683 & 687,487 & 6.875 & 0.000349  \\ 
        \textbf{Arts} & 43,583 & 58,900 & 333,693 & 7.656 & 0.000130  \\ 
        \textbf{Office} & 71,865 & 91,527 & 667,461 & 9.288 & 0.000101  \\
        \textbf{H\&M} & 50,000 & 61,042 & 606,922 & 12.138 & 0.000199  \\ 
    \bottomrule
    \end{tabular}
\end{table}

\section{Experiments}

\subsection{Datasets}
We evaluate the performance of each method by using \textbf{four} datasets, Stream, Arts, Office, and H\&M, which are obtained from \textbf{three} different platforms. The Stream dataset is a stream media dataset from the a video content platform that we have crawled by ourselves. Arts and Office are e-commerce datasets from the Amazon platform\footnote{https://jmcauley.ucsd.edu/data/amazon/}, which are publicly available and commonly used \cite{hou2022learning, zhou2022bootstrap, zhang2022latent}. Arts dataset corresponds to the "Arts, Crafts and Sewing" category of Amazon review datasets, while Office represents the "Office Products" category. H\&M is another e-commerce dataset from the H\&M platform, which is a public competition dataset provided by Kaggle\footnote{https://www.kaggle.com/competitions/h-and-m-personalized-fashion-recommendations}. The diversity of datasets from different platforms helps to demonstrate the robustness of our proposed methods.

For the Stream dataset \footnote{This dataset is provided by the unpublished work \cite{videolens}. If you require access to the relevant data, please refer to Acknowledgement section and contact the authors directly.}, we utilize the video cover and the concatenation of video tags and video titles to represent visual and textual information, respectively. For the other three datasets (Arts, Office, and H\&M), we use the cover of the product to represent visual information. As for textual information, Arts and Office datasets use the concatenation of “title”, “brand”, “category”, and “description”. Different from them, the H\&M dataset consists of the concatenation of “prod\_name”, “product\_type\_name”, “product\_group\_name”, “graphical\_appearance\_name”, and “colour\_group\_name”. 

For all datasets, we utilize user sequences with more than 5 interactions and items that have completely matched textual and visual content. Additionally, we keep only the most recent 15 interaction records for each user. Table ~\ref{exp:dataset} presents the relevant statistical information of each dataset.

\begin{table*}[t]
    \centering
      \renewcommand\arraystretch{1}
    \caption{Overall performance of our model and the baselines on four multi-modal recommendation datasets. Best performances are noted in \textbf{bold}, and the second-best are underlined.}
    \vspace{-0.3cm}
    \begin{tabular}{cl*{8}{c}r}
    \toprule
        \multirow{2}*{\textbf{Dataset}} & \multirow{2}*{\textbf{Metrics}} & \multicolumn{3}{c}{\textbf{General ID-based Sequential Model}} & \multicolumn{5}{c}{\textbf{Sequential Model with Modality Feature}} & \multirow{2}*{\textbf{Improv.}}  \\
        \cmidrule(lr){3-5}\cmidrule(lr){6-10}
        ~ & ~ & \textbf{GRU4Rec} & \textbf{SASRec} & \textbf{NextItNet} & \textbf{FDSA} & \textbf{UniSRec} & \textbf{SASRec+EF} & \textbf{SASRec+LF} & \textbf{ODMT} &   \\ 
        \midrule
        \multirow{4}*{\textbf{Stream}} & Recall@10 & 0.0914  & 0.0935  & 0.0845  & 0.0885  & 0.1067  & \underline{0.1112}  & 0.1090  &\textbf{0.1194}  & \textbf{7.373\%} \\ 
        ~ & NDCG@10 & 0.0492  & 0.0507  & 0.0454  & 0.0476  & 0.0585  & \underline{0.0611}  & 0.0607  & \textbf{0.0672}  & \textbf{9.946\%} \\
        ~ & Recall@20 & 0.1323  & 0.1344  & 0.1253  & 0.1287  & 0.1506  & \underline{0.1572}  & 0.1545  & \textbf{0.1668}  & \textbf{6.132\%} \\
        ~ & NDCG@20 & 0.0595  & 0.0610  & 0.0557  & 0.0577  & 0.0696  & \underline{0.0727}  & 0.0721  & \textbf{0.0791}  & \textbf{8.863\%} \\
        \midrule
        \multirow{4}*{\textbf{Arts}} & Recall@10 & 0.0535  & 0.0617  & 0.0510  & 0.0640  & 0.0860  & 0.0962  & \underline{0.0992}  & \textbf{0.1127}  & \textbf{13.552\%}\\
        ~ & NDCG@10 & 0.0380  & 0.0454  & 0.0363  & 0.0471  & 0.0612  & 0.0669  & \underline{0.0709}  & \textbf{0.0787}  & \textbf{11.127\%} \\
        ~ & Recall@20 & 0.0703  & 0.0787  & 0.0661  & 0.0809  & 0.1095  & 0.1241  & \underline{0.1264}  & \textbf{0.1410}  & \textbf{11.601\%} \\
        ~ & NDCG@20 & 0.0422  & 0.0497  & 0.0401  & 0.0514  & 0.0672  & 0.0739  & \underline{0.0765}  & \textbf{0.0852}  & \textbf{11.308\%} \\
        \midrule
        \multirow{4}*{\textbf{Office}} & Recall@10 & 0.0703  & 0.0769  & 0.0710  & 0.0816  & 0.0971  & 0.1068  & \underline{0.1085}  & \textbf{0.1175}  & \textbf{8.299\%} \\
        ~ & NDCG@10 & 0.0542  & 0.0606  & 0.0533  & 0.0635  & 0.0758  & 0.0814  & \underline{0.0830}  & \textbf{0.0893}  & \textbf{7.558\%} \\
        ~ & Recall@20 & 0.0844  & 0.0914  & 0.0852  & 0.0970  & 0.1154  & 0.1281  & \underline{0.1302}  & \textbf{0.1408}  & \textbf{8.200\%} \\
        ~ & NDCG@20 & 0.0578  & 0.0642  & 0.0568  & 0.0674  & 0.0804  & 0.0867  & \underline{0.0885}  & \textbf{0.0952}  & \textbf{7.575\%} \\
        \midrule
        \multirow{4}*{\textbf{H\&M}} & Recall@10 & 0.0380  & 0.0502  & 0.0275  & 0.0571  & 0.0877  & 0.1115  & \underline{0.1138}  & \textbf{0.1235}  & \textbf{8.522\%} \\ 
        ~ & NDCG@10 & 0.0206  & 0.0296  & 0.0141  & 0.0334  & 0.0531  & 0.0682  & \underline{0.0701}  & \textbf{0.0771}  & \textbf{9.988\%} \\
        ~ & Recall@20 & 0.0582  & 0.0705  & 0.0435  & 0.0830  & 0.1209  & 0.1490  & \underline{0.1513}  & \textbf{0.1629}  & \textbf{7.679\%} \\
        ~ & NDCG@20 & 0.0257  & 0.0347  & 0.0181  & 0.0400  & 0.0615  & 0.0777  & \underline{0.0796}  & \textbf{0.0870}  & \textbf{9.368\%} \\
        \bottomrule
    \end{tabular}
    \label{exp:performance}
\end{table*}

\subsection{Evaluation Metric}
To evaluate the performance of each model, we follow \cite{hou2022towards, rashed2022carca} and adopt the commonly used metrics, Recall@k and NDCG@k (Normalized Discounted Cumulative Gain@k). We report the average results over all users in both metrics, and the higher value indicates better performance. Following \cite{hou2022towards, rashed2022carca}, we use the last interaction as the prediction, the second-to-last as validation, and the rest for training. We conduct hyper-parameter optimization on the validation set, and choose the combination of parameters that yields the highest Recall@10 as the optimal configuration.

\subsection{Implementation Details}
To make a fair comparison, we reproduce all of the baselines by utilizing our pipeline framework. Our default loss function for all models is debiased in-batch cross-entropy loss. The visual features are extracted based on the "openai/clip-vit-base-patch32" \cite{radford2021learning} pre-trained model. The textual features with Chinese words are extracted by using the "hfl/chinese-roberta-wwm-ext" \cite{cui-etal-2020-revisiting} pre-trained model and textual features with English words are extracted by using the "bert-base-uncased" \cite{devlin2018bert} pre-trained model. 
We conduct the grid search for hyper-parameters, such as hidden size and learning rate. For the general sequential models, FDSA \cite{zhang2018deep} and UniSRec \cite{hou2022towards}, the search range for hidden sizes and learning rates are [128, 256, 512, 768] and [1e-3, 1e-4, 1e-5], respectively. Empirically, SASRec+EF and SASRec+LF both have two Transformer layers for text and image modalities, culminating in a total of four Transformer layers. In our model, we adopt two Transformer layers of IMT. 
In our experiments, we set the batch size to 128, which includes 128 sequences from different users per batch for training. For specific hyper-parameters unique to each baseline, such as the number of GRU layers in GRU4Rec and the selection of dilated convolution layers in NextItNet, we refer to the settings in RecBole \cite{zhao2021recbole}. 

\subsection{Comparison with SOTA Methods}
Based on the input types, we divide SOTA methods into two categories: 1) \textbf{general sequential recommendation models} that only take item ID information as input (e.g. GRU4Rec \cite{hidasi2015session}, SASRec \cite{kang2018self}, NextItNet \cite{yuan2019simple}), and 2) \textbf{multi-modal sequential recommendation models} that take both item ID information and modality information (visual and textual) as input (e.g. FDSA \cite{zhang2019feature}, UniSRec \cite{hou2022towards}, SASRec+EF, SASRec+LF): \textbf{(a). GRU4Rec} is a session-based recommendation algorithm that uses recurrent neural networks (GRUs) to model user behavior; \textbf{(b). SASRec} is a self-attention-based sequential recommendation algorithm that uses a multi-head self-attention mechanism to capture user preferences; \textbf{(c). NextItNet} is a neural network-based sequential recommendation algorithm that uses dilated convolutions to capture long-term dependencies between items; \textbf{(d). FDSA} is a feature-driven and self-attention-based sequential recommendation algorithm that uses feature-driven attention mechanisms to capture user preferences; \textbf{(e). UniSRec} is a universal sequence representation learning algorithm for recommendation that harnesses the descriptive text associated with an item to learn transferable representations across different domains and platforms; \textbf{(f). SASRec+EF \ (Our Extension)} is an extension of SASRec that takes id, text, and image as input and uses Transformer layers as the FST module with Early Fusion; \textbf{(g). SASRec+LF \ (Our Extension)} is an extension of SASRec that takes id, text, and image as input and uses Transformer layers as the FST module with Late Fusion.

For the vanilla UniSRec \cite{hou2022towards} and FDSA \cite{zhang2018deep} models, only textual information is utilized. To make a fair comparison, we reproduce these models by incorporating image information, leveraging the inherent extensibility of the UniSRec and FDSA models.

Table ~\ref{exp:performance} demonstrates the superiority of the \textbf{Sequential Model with Modality Feature} over the \textbf{General ID-based Sequential Model} across all datasets and evaluation metrics, which underscores the potential benefits of incorporating modality information to enhance recommendation accuracy. Notably, SASRec+EF and SASRec+LF within the \textbf{Sequential Model with Modality Feature} outperform UniSRec and FDSA in terms of Recall@10 and NDCG@10 scores, indicating that utilizing Transformers as FST module may lead to more effective item representation modeling and improves recommendation accuracy. Furthermore, the fusion strategy of Late Fusion is proven to result in better overall performance. Based on the SASRec+LF framework, our proposed ODMT model aims to achieve multi-source information representation learning in a unified manner, which leverages the strengths of contemporary multi-modal Transformer models and online distillation methods. Experimental results demonstrate that our proposed ODMT model can achieve better performance across all four datasets and four metrics, surpassing not only SASRec+LF but also all other baseline models.

\begin{table}[b]
    \centering
      \renewcommand\arraystretch{1}
    \caption{Ablation analysis results on two downstream datasets. "R@10" is short
for Recall@10, and "N@10" is short for NDCG@10. }
\vspace{-0.4cm}
    \label{exp:ablation}
    \begin{tabular}{l*{4}{c}}
    \toprule
        \textbf{Dataset} & \multicolumn{2}{c}{\textbf{Stream}} & \multicolumn{2}{c}{\textbf{Arts}}\\
        \cmidrule(lr){1-1} \cmidrule(lr){2-3} \cmidrule(lr){4-5}
        \textbf{Method} & \textbf{R@10} & \textbf{N@10} & \textbf{R@10} & \textbf{N@10} \\ 
        \toprule
        \textbf{Text Initialization} & 0.1191  & 0.0666  & 0.1107  & 0.0777  \\
        \textbf{Image Initialization} & 0.1161  & 0.0647  & 0.1051  & 0.0738  \\
        \textbf{w/o Initialization} & 0.1144  & 0.0638  & 0.1025  & 0.0722  \\ 
        \textbf{w/o ID mask} & 0.1155  & 0.0646  & 0.1001  & 0.0700  \\ 
        \textbf{w/o IMT (1)} & 0.1088  & 0.0607  & 0.1068  & 0.0752  \\ 
        \textbf{w/o IMT (2)} & 0.1146  & 0.0636  & 0.1096  & 0.0770  \\ 
        \textbf{w/o Online Distillation} & 0.1121  & 0.0626  & 0.1019  & 0.0723  \\ 
        \textbf{w/o ID} & 0.1125  & 0.0623  & 0.1075  & 0.0747  \\ \midrule
        \textbf{ODMT (full framework)} & \textbf{0.1194}  & \textbf{0.0672}  & \textbf{0.1127}  & \textbf{0.0787} \\ 
    \bottomrule
    \end{tabular}
\end{table}

\subsection{Ablation Study}

In this study, we conduct an analysis to evaluate the impact of each module on the final performance of our proposed ODMT model. To compare the performance of our model with other variants, we prepare \textbf{8} different models, including:

\textbf{(1) Text Initialization}, which initializes the ID embedding table using only textual features;
\textbf{(2) Image Initialization}, which initializes the ID embedding table using only visual features;
\textbf{(3) w/o Initialization}, which abandons initialization for the ID embedding table with text and image features;
\textbf{(4) w/o ID mask}, which removes the limitation of modality features being unable to attend to ID features in the IMT module;
\textbf{(5) w/o IMT (1)}, which replaces the two layers of IMT with two standard Transformer layers for text input and two  standard Transformer layers for image input, with the depth of Transformers in both cases remaining the same;
\textbf{(6) w/o IMT (2)}, which replaces the two layers of IMT with one standard Transformer layer for text input and one standard Transformer layer for image input, while keeping the total number of Transformers the same;
\textbf{(7) w/o Online Distillation}, which uses a traditional Late Fusion loss calculation method and removes the distillation loss;
\textbf{(8) w/o ID}, which removes the ID component in ODMT and replaces it with the corresponding ID-free version.

As shown in Table \ref{exp:ablation}, each new component contributes to the final performance. In the ID initialization part, utilizing average features of both text and image modalities for initializing the ID embedding table results in the best performance, with text features following closely. However, random initialization of the ID embedding table has a detrimental impact on prediction results, particularly affecting the optimization of the IMT module. "w/o ID" shows an obvious performance reduction compared to full ODMT, which highlights the effective accommodation of both ID features and multi-modal features in our framework. 

\vspace{-0.2cm}
\subsection{In-depth Analysis}
\subsubsection{Performance Comparison w.r.t Item Popularity}
\begin{figure}[t]
    \begin{minipage}{0.5\columnwidth}
      \includegraphics[width=\linewidth]{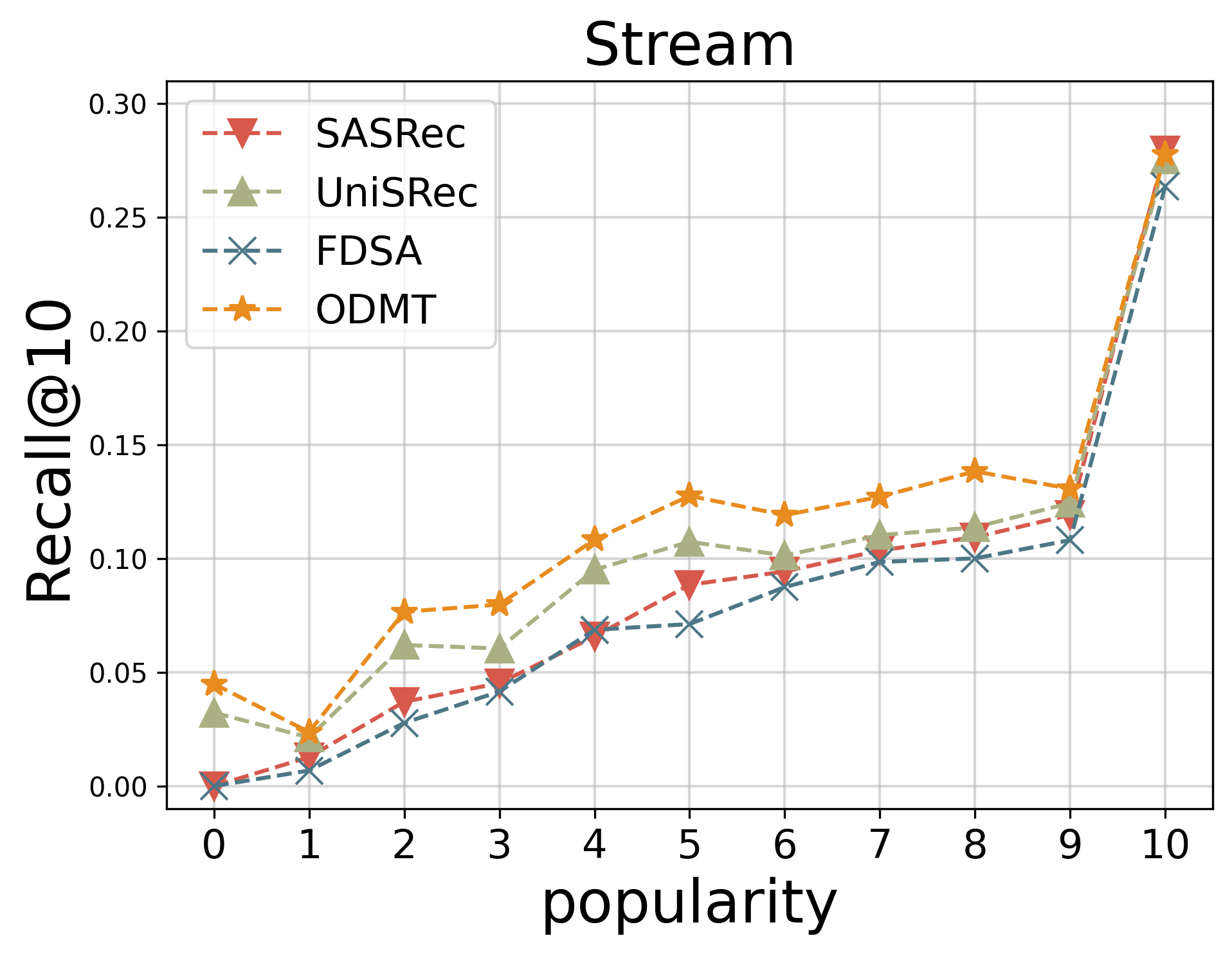}
    \end{minipage}\hfill 
    \begin{minipage}{0.5\columnwidth}
      \includegraphics[width=\linewidth]{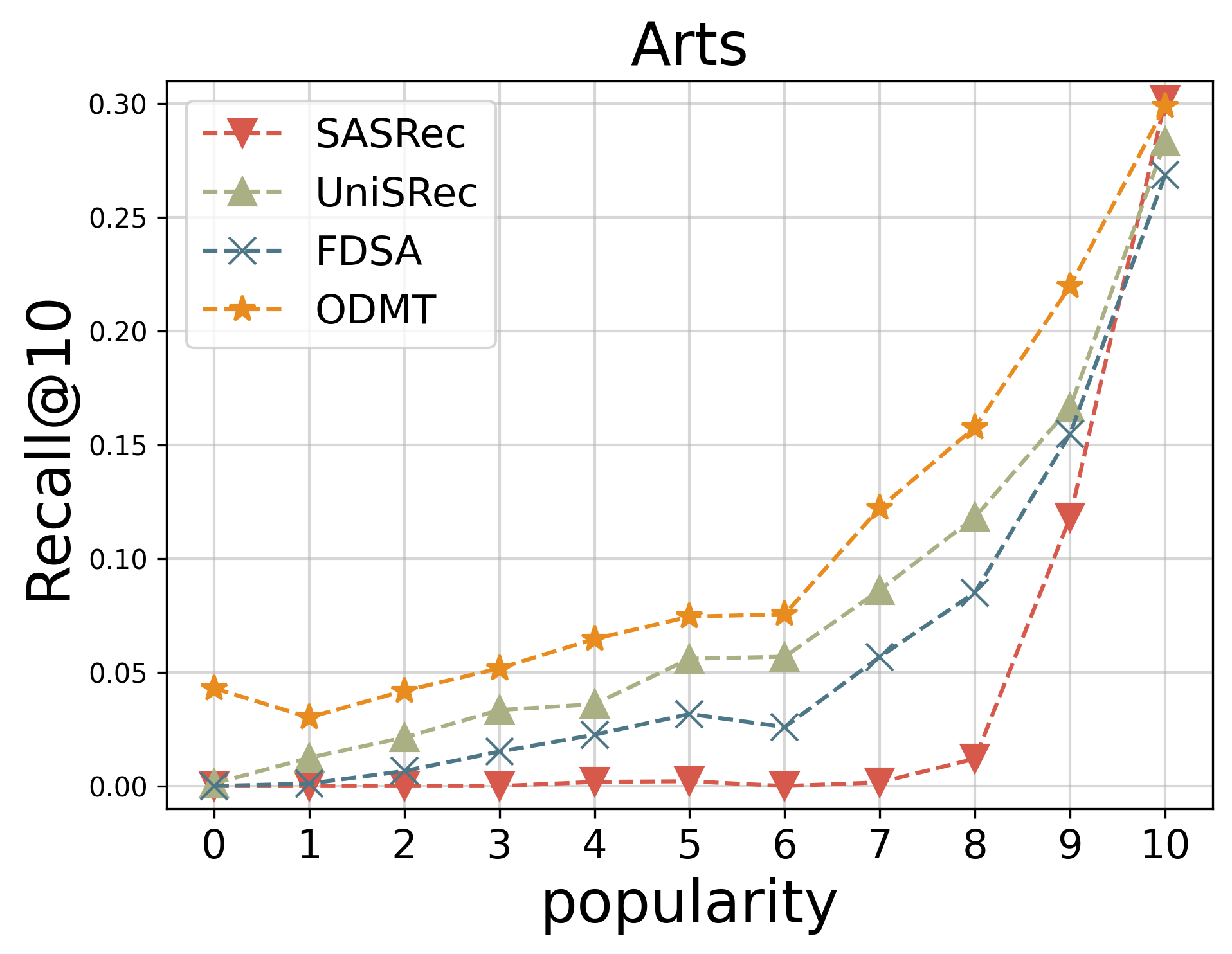}
    \end{minipage}
    \vspace{-0.4cm}
    \caption{Performance comparison of sequential models regarding item popularity, where the x-axis represents different groups divided based on quantile statistics. Group 0 refers to items that have not appeared in the training set, while the item count is kept consistent across other groups. }
    \label{exp:popularity}
\end{figure}

The item popularity distribution follows a Matthew effect, with a majority of users showing interest in only a small portion of items. Figure \ref{exp:popularity} indicates existing models predict popular items well but struggle with long-tail items. Moreover, Figure \ref{exp:popularity} illustrates the consistent improvement of our proposed method compared to the SOTA models in item groups with varying popularity. Notably, in item group 0, which consists of items not present in the training set, the ID-based SASRec fails to make accurate predictions. In contrast, our proposed model demonstrates effective mitigation of the cold-start problem, showcasing superior performance compared to other multi-modal sequential recommendation models.

\subsubsection{Different Sequential Models as Backbone.}
\begin{figure}[t]
    \begin{minipage}{0.5\columnwidth}
      \includegraphics[width=\linewidth]{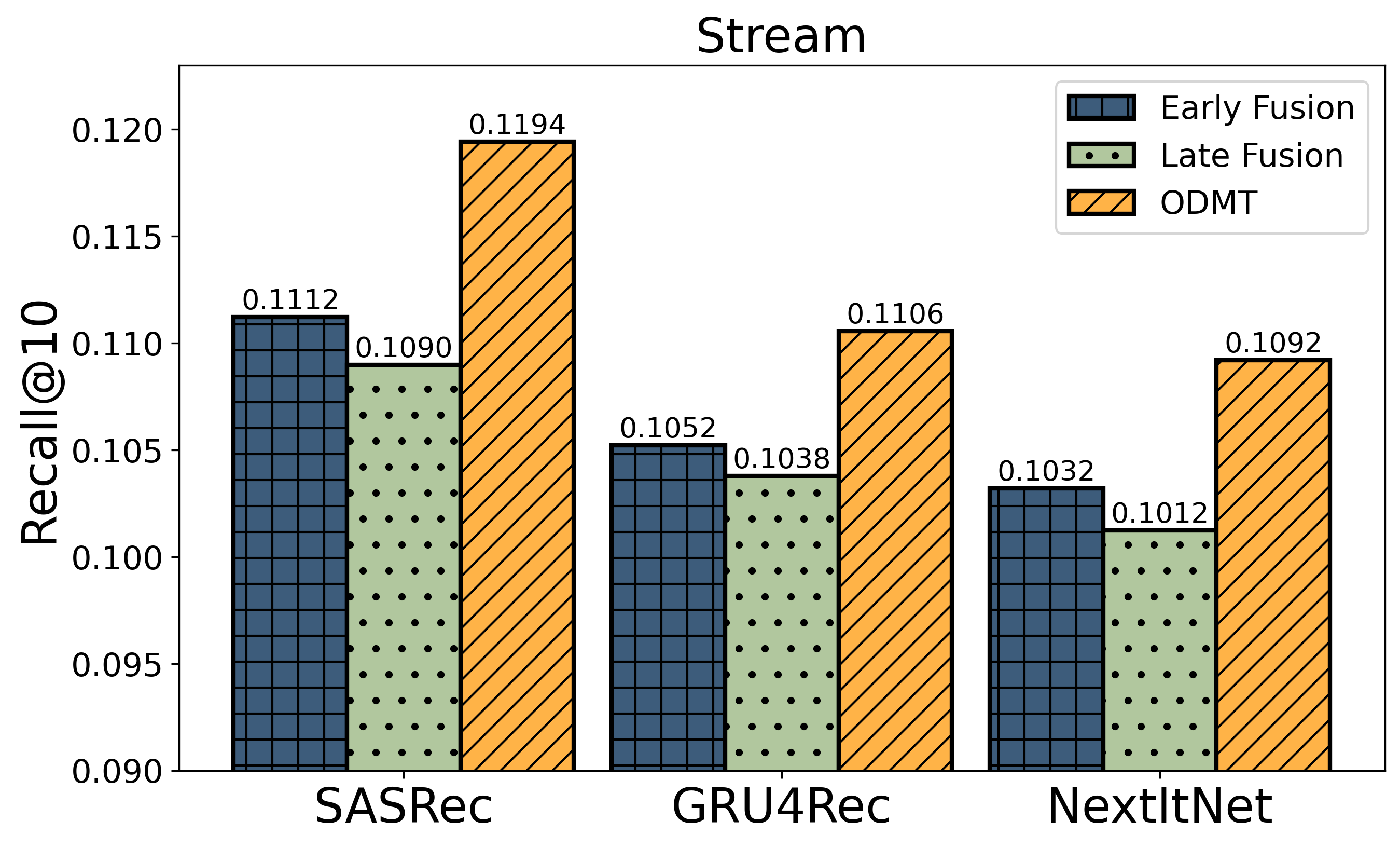}
    \end{minipage}\hfill 
    \begin{minipage}{0.5\columnwidth}
      \includegraphics[width=\linewidth]{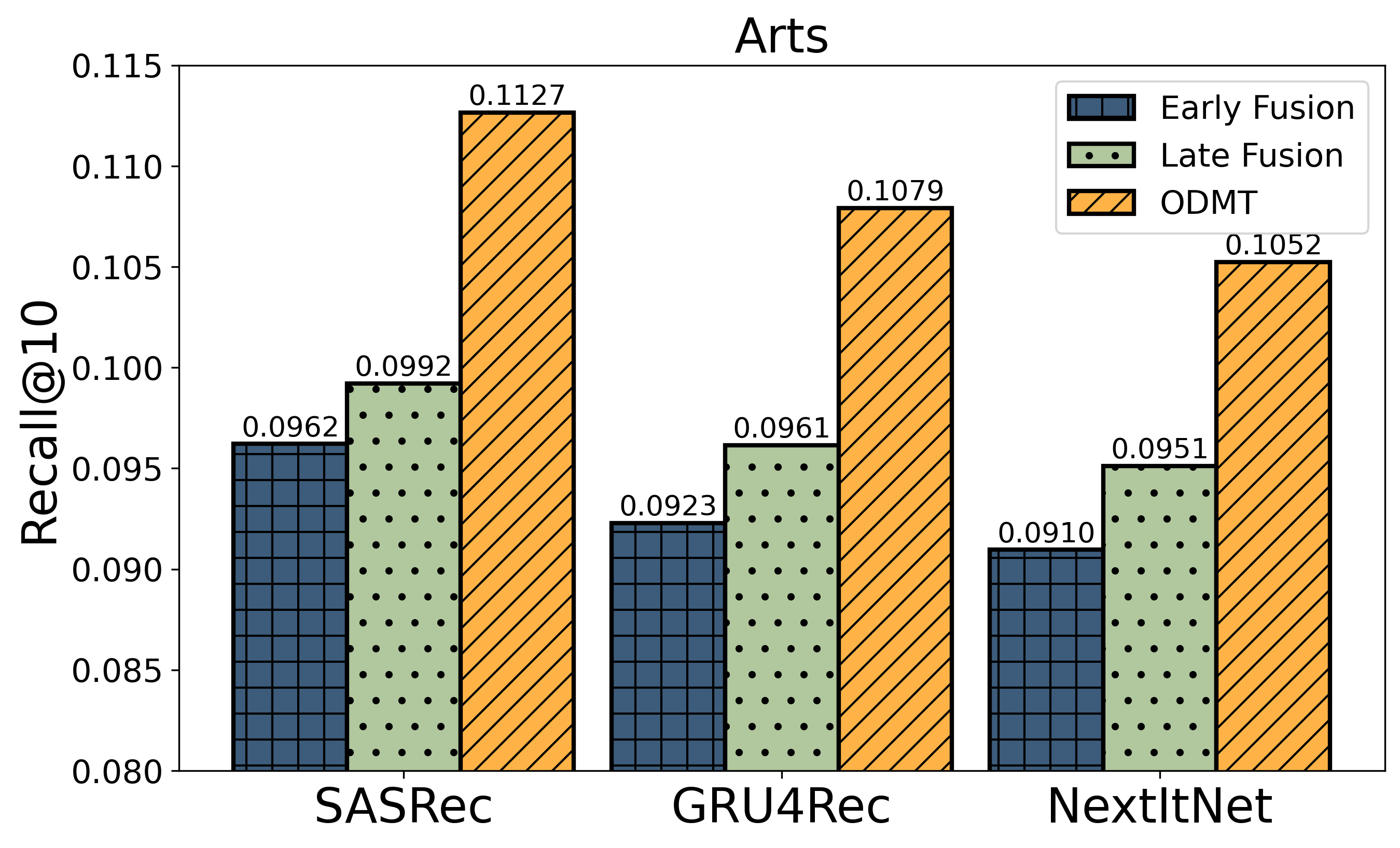}
    \end{minipage}
    \vspace{-0.3cm}
    \caption{Performance comparison of sequential models with different backbones.}
    \label{exp:backbone}
\end{figure}

Since our method is model-agnostic, we conduct experiments based on various sequential models to test the robustness. Figure \ref{exp:backbone} shows that ODMT can surpass other methods consistently. This highlights the generality of ODMT, which can be seamlessly integrated as a plug-and-play module into any general sequential model.

\section{Related Work}
\subsection{Multi-modal Recommendation}
The development of computer vision~\cite{ji2020context}, natural language processing~\cite{li2021mrn}, and multi-modal learning~\cite{zhang2023transfer,ji2023partial,ji2023binary,ji2022mrtnet} has provided better representations for heterogeneous data structures. Recently, multi-modal representations have been widely used in the field of recommendation systems, where collaborative filtering paradigms still dominate, with multi-modal features typically incorporated as side information in the model framework \cite{wei2019mmgcn, wei2020graph, zhou2022bootstrap,DROS,wei2023lightgt}. In the field of sequential recommendation, several studies have found that multi-modal features can also yield significant improvements \cite{hou2022towards, hou2022learning, rashed2022carca}. Especially, \cite{yuan2023go} even achieved comparable results to traditional ID features using only multi-modal features, which underscores the significance of modal features in sequential recommendation and their substitutability for traditional ID features. We attribute this to the characteristics of sequential recommendation models, which heavily rely on item representations for modeling user preferences, unlike collaborative filtering which requires additional learning of user representations. Through extensive experimental exploration of multi-modal sequential recommendation, we observe the issue of conflicts between multi-modal and ID features during training. Based on these observations, we design two modules that leverage the characteristics of ID and multi-modal features to make them compatible and mutually beneficial.

\vspace{-0.2cm}
\subsection{Knowledge Distillation}
Knowledge distillation \cite{hinton2015distilling,yin2022content} aims to guide the learning of a student model by using a pretrained teacher model, allowing the student model to achieve better predictive performance with smaller model sizes. In recent years, online distillation has gained attention due to its end-to-end training strategy, which eliminates the need for a pre-trained teacher model. Unlike the traditional "teacher-student" paradigm of knowledge distillation, online distillation allows for mutual learning between all sub-networks \cite{anil2018large, zhang2018deep} or the use of ensemble methods to obtain a teacher output that combines multiple prediction results \cite{zhu2018knowledge, guo2020online, kim2021feature}, which in turn guide the learning of all sub-networks. Online distillation typically requires that each sub-network can independently complete downstream prediction tasks (usually classification tasks) and that the prediction results of different sub-networks have diverse characteristics \cite{chen2020online}. In sequential recommendation, we find that using ID features or multi-modal features (such as image or text) can independently complete recommendation predictions. Additionally, due to the heterogeneity of the input, the outputs of different features are also diverse. Based on these findings, we propose a multi-modal online distillation framework for sequential recommendation models.

\vspace{-0.1cm}
\section{Conclusion}
This paper investigates the impact of item representation learning on downstream recommendation tasks, exploring disparities in information fusion at different stages. Empirical experiments show significant influence on recommendation performance. To enhance recommendation accuracy, the paper proposes two novel modules: ID-aware Multi-modal Transformer for feature interaction and online distillation training for multi-faceted user interest distribution and improved prediction robustness. Experimental results on four datasets demonstrate the effectiveness of these modules.

\section{Acknowledgement}
This work is supported by the Advanced Research and Technology Innovation Centre (ARTIC), the National University of Singapore under Grant (project number: A-8000969-00-00). This research is also supported by the National Natural Science Foundation of China (9227010114) and the University Synergy Innovation Program of Anhui Province (GXXT-2022-040). Futhermore, we extend our gratitude to the Lab for Representation Learning at Westlake University (fajieyuan@westlake.edu.cn) for supporting dataset.

{\small
\bibliographystyle{ACM-Reference-Format}
\balance
\bibliography{recommendation
}
}

\clearpage

\appendix
\section{Appendix}
\subsection{Dataset Details}
\begin{figure}[b]
  \centering
  \includegraphics[width=.5\linewidth - 0.25mm]{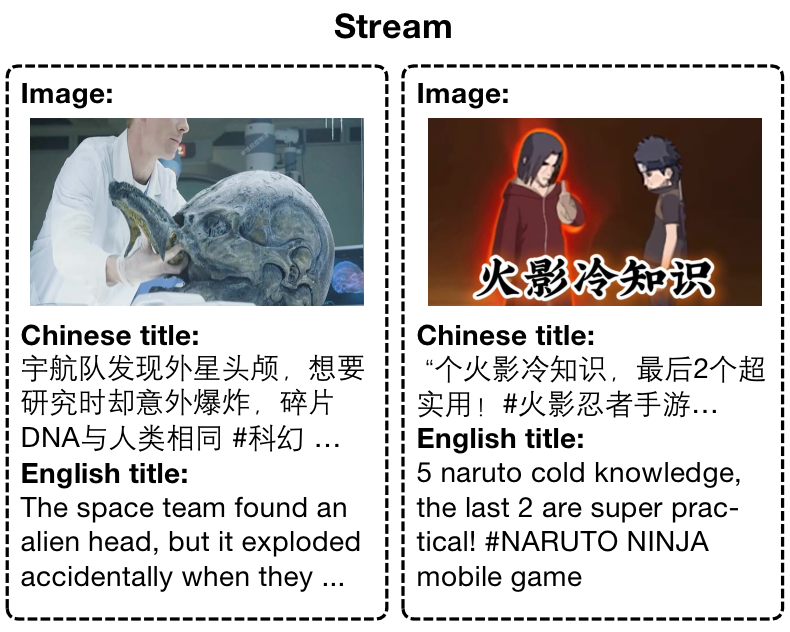}\hfill
  \includegraphics[width=.5\linewidth - 0.25mm]{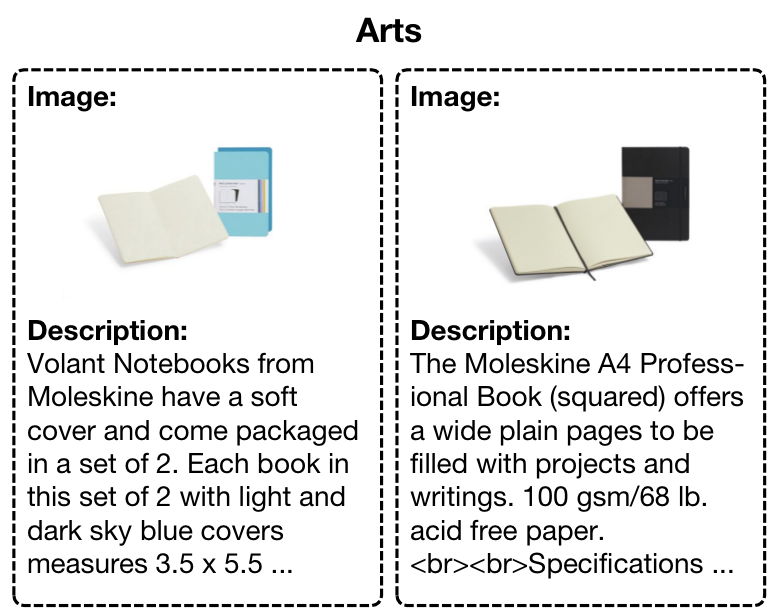}\\[0.5mm]
  \includegraphics[width=.5\linewidth - 0.25mm]{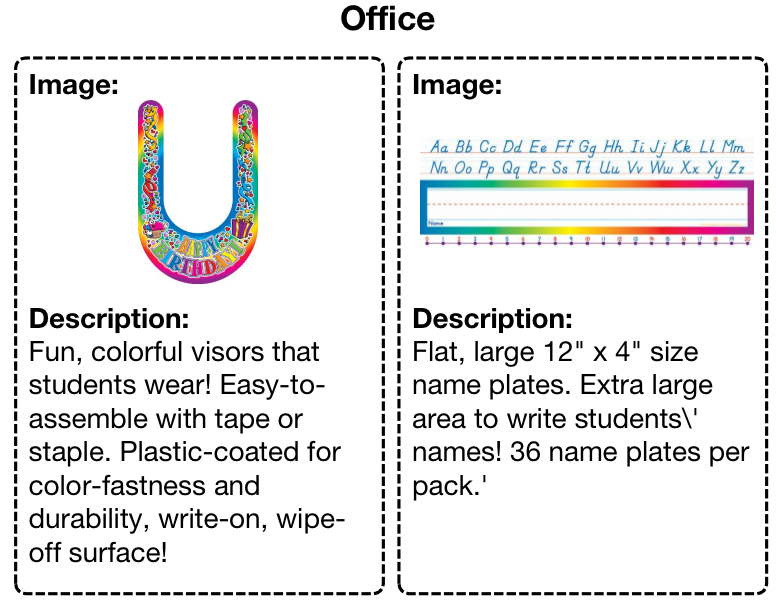}\hfill
  \includegraphics[width=.5\linewidth - 0.25mm]{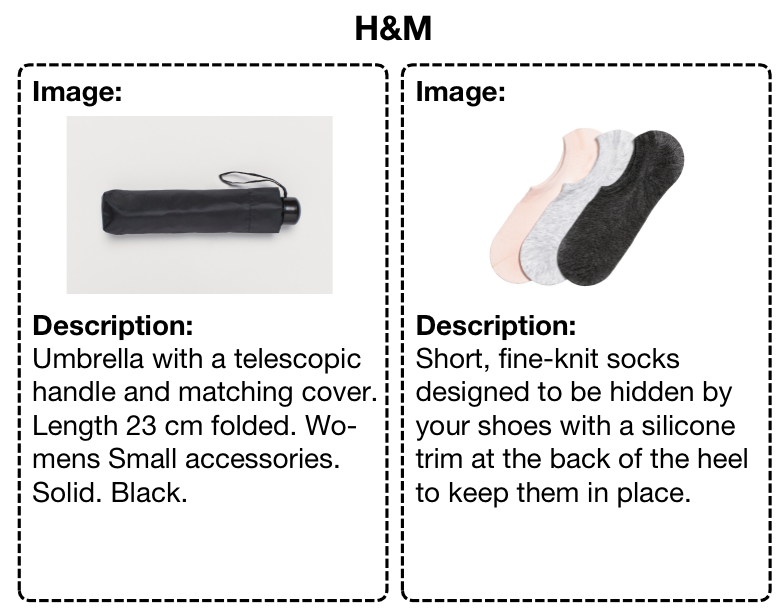}
\caption{Selected sample cases from four different datasets.}
\label{exp:case}
\end{figure}

Figure ~\ref{exp:case} shows some samples of \textbf{four} datasets from \textbf{three} platforms, namely Stream, Amazon and H\&M, respectively. In contrast to the majority of existing research that predominantly focuses on experiments conducted on Amazon datasets \cite{hou2022towards, hou2022learning, rashed2022carca, liu2022disentangled}, our study employs a diverse range of experimental datasets with the aim of validating the robustness of our proposed method.

As for the Stream dataset, it is assembled by gathering publicly accessible user comments on videos from the stream media platform between February 2020 and September 2022. Specifically, we first collect short videos from the homepage to ensure the channel diversity. Subsequently, more videos are included via crawling related videos from the associated page of each video during the first stage. Finally, we establish the user set by acquiring publisher IDs from the comment sections under all videos. Note that there are no privacy issues since all videos and user comments are publicly available.

\subsection{Effect of Different Transformer Layers} 
\begin{figure}[b]
    \begin{minipage}{0.5\columnwidth}
      \includegraphics[width=\linewidth]{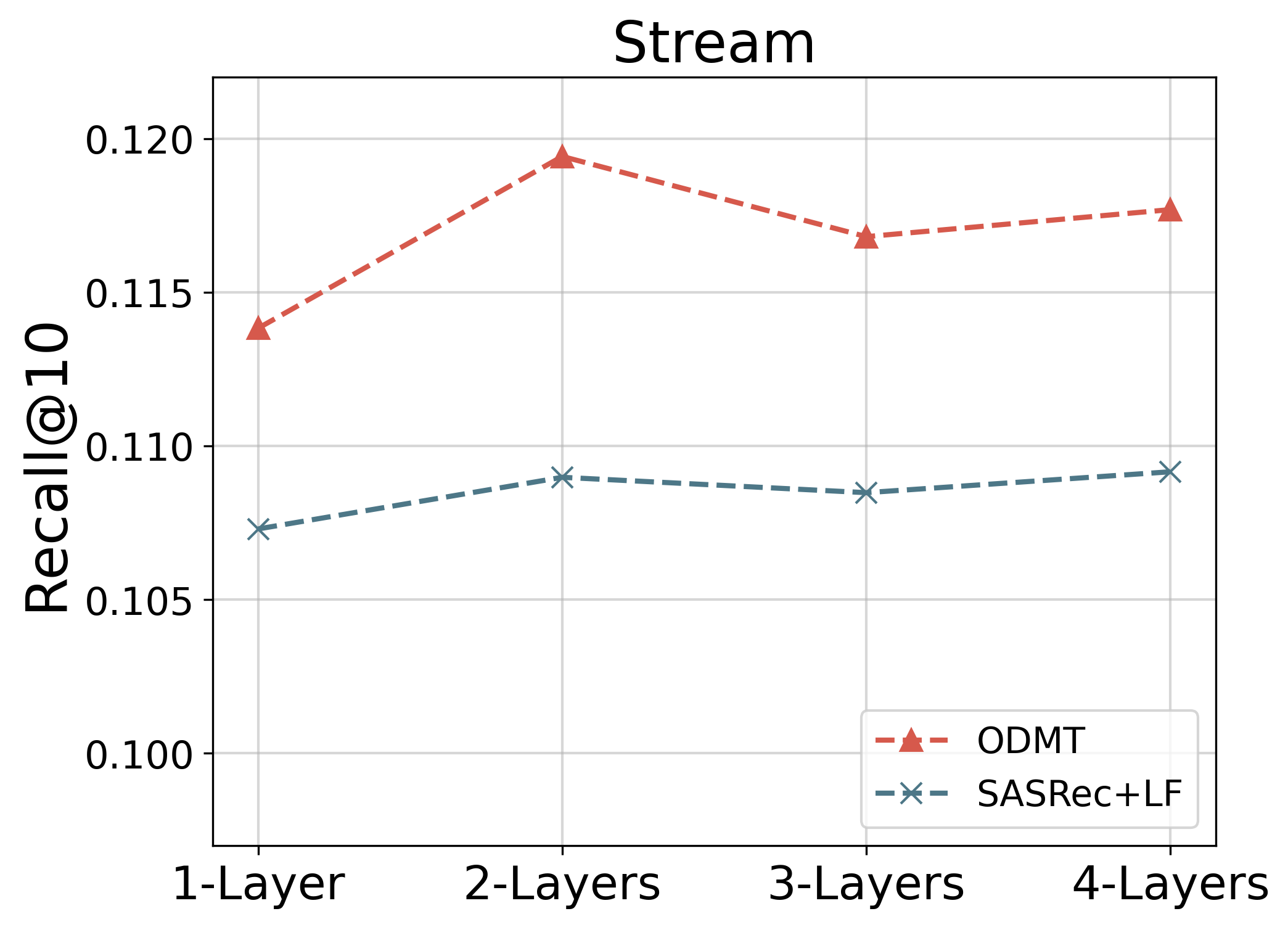}
    \end{minipage}\hfill 
    \begin{minipage}{0.5\columnwidth}
      \includegraphics[width=\linewidth]{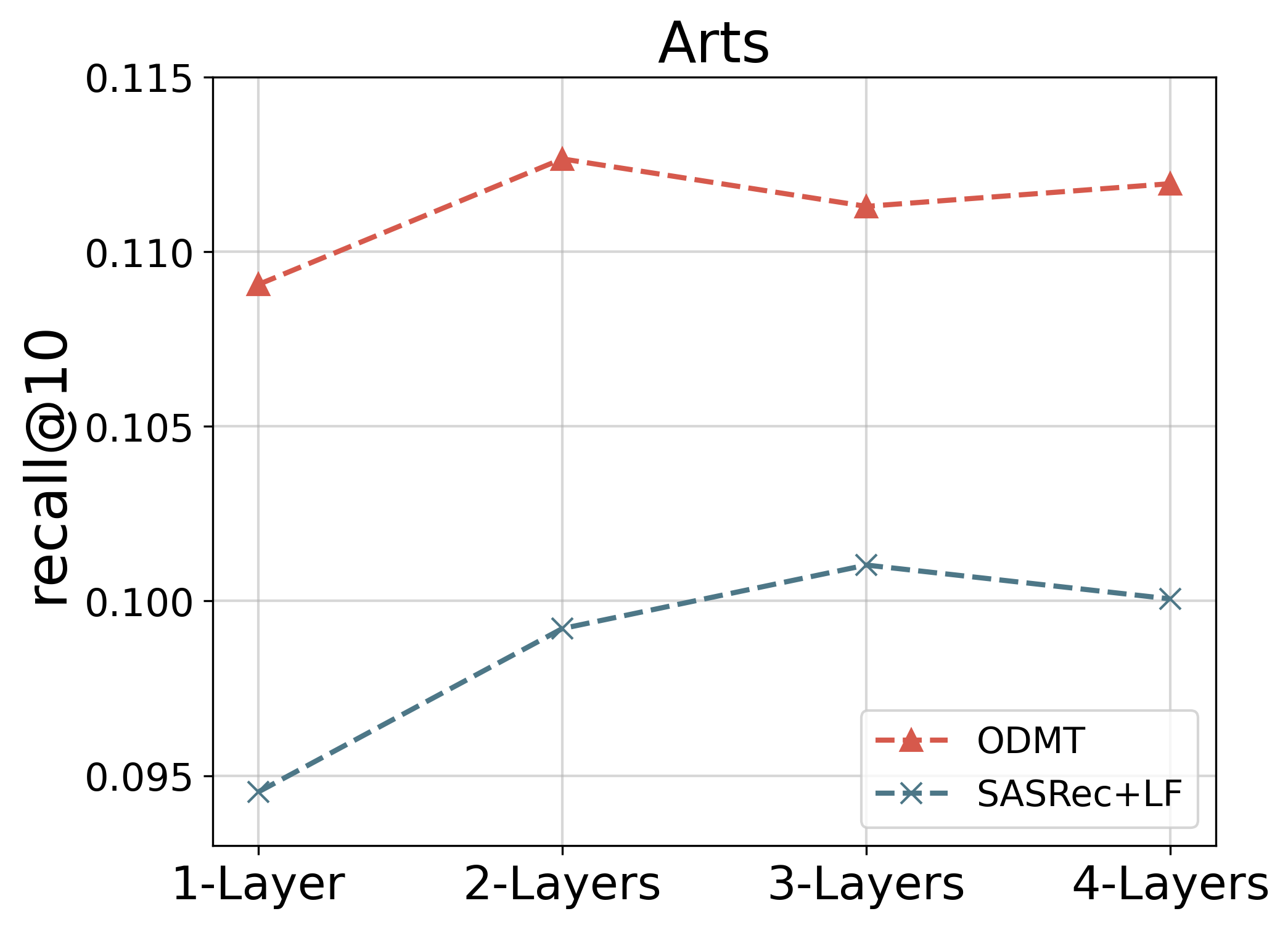}
    \end{minipage}
    \caption{Performance comparison of different numbers of Transformer layers.}
    \label{exp:layers}
\end{figure}

We investigate the number of layers in the multi-modal Transformer of the IMT module, where the number of multi-heads corresponds to the number of Transformer layers. For comparison, we also explore the SASRec+LF method under the same conditions. Figure \ref{exp:layers} indicates that 2 Transformer layers are sufficient to achieve good performance, which is our default setting. Comparing our model with SASRec+LF, ODMT consistently outperforms SASRec+LF across different numbers of Transformer layers.

\subsection{Effect of Parameters in Knowledge Distillation}
The online distillation part of our framework involves two important parameters: 1) \textbf{Temperature parameter (T)} serves as a scaling factor that controls the softness or hardness of predicted probabilities. It affects the balance between exploration and exploitation during the distillation process by controlling the level of smoothing or sharpening in the probability distributions. We set $\rm{T}=0.5$ on the Stream dataset, $\rm{T}=0.3$ on the Arts dataset, $\rm{T}=0.4$ on the Office dataset and $\rm{T}=0.5$ on the H\&M dataset. 
2) \textbf{Weight factor (\bm{$\alpha$})} controls the weight of the distillation loss in the whole loss function. A larger value of this parameter results in a smaller weight for the distillation loss at a fixed epoch, and the weight increases exponentially with the epoch. This allows the model to prioritize learning from the distillation loss during the late stage of training, mitigating error accumulation from poor student model performance in the early stages. We set $\alpha=50$ on the Stream dataset, $\alpha=20$ on the Arts dataset, $\alpha=40$ on the Office dataset, and $\alpha=60$ on the H\&M dataset.
Table ~\ref{exp:knowledge} illustrates the impact of the two parameters on the experimental results on Stream and Arts datasets.

\begin{table}[h]
    \centering
    \caption{Performance comparison of different hyper-parameters settings in the online distillation module. "None" denotes the weight of distillation loss is 0, primarily used for control purposes in comparison with other parameters. "R@10" is short for Recall@10, "N@10" is short for NDCG@10. The best performance is highlighted in bold.}\label{exp:knowledge}
    \begin{tabular}{|c|c|c|c|c|c|}
    \hline
        \multicolumn{6}{|c|}{Stream} \\ \hline
        T ($\alpha$=50) & R@10 & N@10 & $\alpha$ (T=0.5) & R@10 & N@10 \\ \hline
        None & 0.1159  & 0.0645  & None & 0.1159  & 0.0645  \\ \hline
        0.1 & 0.1162  & 0.0648  & 10 & 0.1177  & 0.0654  \\ \hline
        0.2 & 0.1159  & 0.0650  & 20 & 0.1177  & 0.0659  \\ \hline
        0.3 & 0.1171  & 0.0652  & 30 & 0.1186  & 0.0662  \\ \hline
        0.4 & 0.1183  & 0.0661  & 40 & 0.1179  & 0.0660  \\ \hline
        \textbf{0.5} & \textbf{0.1194}  & \textbf{0.0672} & \textbf{50} & \textbf{0.1194}  & \textbf{0.0672}  \\ \hline
        0.6 & 0.1194  & 0.0669  & 60 & 0.1178  & 0.0657  \\ \hline
        \multicolumn{6}{|c|}{Arts} \\ \hline
        T ($\alpha$=20) & R@10 & N@10 & $\alpha$ (T=0.3) & R@10 & N@10 \\ \hline
        None & 0.1091  & 0.0766  & None & 0.1091  & 0.0766  \\ \hline
        0.1 & 0.1120  & 0.0784  & 10 & 0.1120  & 0.0783  \\ \hline
        0.2 & 0.1105  & 0.0771  & \textbf{20} & \textbf{0.1127}  & \textbf{0.0787}  \\ \hline
        \textbf{0.3} & \textbf{0.1127}  & \textbf{0.0787}  & 30 & 0.1113  & 0.0777  \\ \hline
        0.4 & 0.1120  & 0.0785  & 40 & 0.1103  & 0.0772  \\ \hline
        0.5 & 0.1091  & 0.0761  & 50 & 0.1118  & 0.0780  \\ \hline
        0.6 & 0.1058  & 0.0738  & 60 & 0.1107  & 0.0776 \\ \hline
    \end{tabular}
\end{table}


\end{document}